\documentclass{emulateapj}
\usepackage{graphicx}
\usepackage{amsfonts}
\usepackage{pifont}

\def\ms{\mbox{$M_*$}}
\def\mh{\mbox{$M_{\rm h}$}}

\def\mhchar{\mbox{$M_{\rm h}^{\star}$}}  
\def\msatb{\mbox{$M_{*,\rm sat}^b$}}
\def\msatj{\mbox{$M_{*,\rm sat}^j$}}

\def\msun{\mbox{M$_{\odot}$}}

\def\Phicen{\mbox{$\Phi_{c,t}(\ms|\mh)$}}
\def\Phicenb{\mbox{$\Phi_{c,b}(\ms|\mh)$}}
\def\Phicenr{\mbox{$\Phi_{c,r}(\ms|\mh)$}}
\def\Pcen{\mbox{$P_{c,t}(\ms|\mh)$}}
\def\Pbc{\mbox{$P_{c,b}(\ms|\mh)$}}
\def\Prc{\mbox{$P_{c,r}(\ms|\mh)$}}

\def\mbc{\mbox{$M_{*,b}$}}
\def\mrc{\mbox{$M_{*,r}$}}

\def\phigc{\mbox{$\phi_{g_c,t}$}}
\def\phigbc{\mbox{$\phi_{g_c,b}$}}
\def\phigrc{\mbox{$\phi_{g_c,r}$}}

\def\phig{\mbox{$\phi_{g,t}$}}

\def\phih{\mbox{$\phi_{h}$}}
\def\phibh{\mbox{$\phi_{h,b}$}}
\def\phirh{\mbox{$\phi_{h,r}$}}

\def\fb{\mbox{$f_{\rm blue}$}}
\def\fr{\mbox{$f_{\rm red}$}}
\def\fj{\mbox{$f_{j}$}}

\def\Ncen{\mbox{$\langle N_{c}(>\ms|\mh)\rangle$}}

\def\Ncjj{\mbox{$\langle N_{c,j}\rangle$}}
\def\Pcj{\mbox{$P_{c,j}(\ms|\mh)$}}
\def\Phicj{\mbox{$\Phi_{c,j}(\ms|\mh)$}}

\def\sigmaj{\mbox{$\sigma_{j}$}}   

\def\Phisat{\mbox{$\Phi_{s,t}(\ms|\mh)$}}

\def\Nsatj{\mbox{$\langle N_{s,j}(>\ms|\mh) \rangle$}}
\def\Nsatjj{\mbox{$\langle N_{s,j}\rangle$}}

\def\fbsat{\mbox{$f_{b,\rm sat}(\ms|\mh)$}}
\def\frsat{\mbox{$f_{r,\rm sat}(\ms|\mh)$}}
\def\fjsat{\mbox{$f_{j,\rm sat}(\ms|\mh)$}}

\def\Phisatj{\mbox{$\Phi_{s,j}(\ms|\mh)$}}

\def\phigs{\mbox{$\phi_{g_{s,t}}$}}
\def\phigbs{\mbox{$\phi_{g_{s,b}}$}}
\def\phigrs{\mbox{$\phi_{g_{s,r}}$}}

\def\phigsij{\mbox{$\phi_{g_{i,j}}$}}
\def\Phiij{\mbox{$\Phi_{i,j}(\ms|\mh)$}}
\def\ngij{\mbox{$n_{g_i,j}$}}
\def\Nsij{\mbox{$\langle N_{i,j}(>\ms|\mh)\rangle$}}
\def\Phiijint{\mbox{$\Phi_{i,j}(M_*'|\mh)$}}

\def\xij{\mbox{$\xi_{gg,j}(r)$}}
\def\xiggj{\mbox{$\xi_{gg,j}^{\rm 1h}(r)$}}
\def\xiggggj{\mbox{$\xi_{gg,j}^{\rm 2h}(r)$}}
\def\Nij{\mbox{$\langle N_{j}(N_{j}-1)\rangle$}}
\def\Nss{\mbox{$\langle N_{s,j}(N_{s,j}-1)\rangle$}}
\def\Navej{\mbox{$\langle N_{j}(\mh)\rangle$}}

\def\Nssij{\mbox{$\langle N_{s,j}\rangle^2$}}
\def\lnfw{\mbox{$\lambda_{h}(r)$}}
\def\lnfwcs{\mbox{$\lambda_{c,s}(r)$}}
\def\lnfwss{\mbox{$\lambda_{s,s}(r)$}}

\def\gsmf{\mbox{GSMF}}
\def\pcf{\mbox{2PCF}}
\def\amt{\mbox{SHAM}}
\def\hod{\mbox{HOD}}
\def\lcdm{\mbox{$\Lambda$CDM}}
\def\shmr{\mbox{SHMR}}
\def\cmf{\mbox{CSMF}}
\def\hmf{\mbox{HMF}}

\defcitealias{Yang+2012}{Y12}
\defcitealias{RAD13}{RAD13}
\defcitealias{RDA12}{RDA12}

\def\nyu{\mbox{NYU-VAGC}}

\def\aap{\mbox{A\&A}}
\def\apjl{\mbox{ApJ}}
\def\apjs{\mbox{ApJS}}

\def\ltsima{$\; \buildrel < \over \sim \;$}    
\def\lesssim{\lower.5ex\hbox{\ltsima}}           
\def\gtsima{$\; \buildrel > \over \sim \;$}    
\def\grtsim{$\lower.5ex\hbox{\gtsima}$}           

\slugcomment{Submitted to ApJ}

\shorttitle{Connecting blue and red galaxies to their host halos}
\shortauthors{Rodr\'iguez-Puebla et al.}

\begin{document}


\title{The stellar-to-halo mass relation of local galaxies segregates by color}  


\author{Aldo Rodr\'iguez-Puebla\altaffilmark{1,2}, Vladimir Avila-Reese\altaffilmark{2}, 
Xiaohu Yang\altaffilmark{1,3}, Sebastien Foucaud\altaffilmark{1}, Niv Drory\altaffilmark{4}, Y. P. Jing\altaffilmark{1}}
\affil{\altaffilmark{1}Center for Astronomy and Astrophysics, Shanghai Jiao Tong University, 
Shanghai 200240, China,
}
\affil{\altaffilmark{2}Instituto de Astronom\'ia, 
Universidad Nacional Aut\'onoma de M\'exico,
A. P. 70-264, 04510, M\'exico, D.F., M\'exico,}
\affil{\altaffilmark{3}Key Laboratory for Research in Galaxies and Cosmology,
Shanghai Astronomical Observatory, Nandan Road 80, Shanghai 200030, China,}
\affil{\altaffilmark{4}McDonald Observatory, The University of Texas at Austin, 1 University Station, Austin, TX 78712-0259, USA.}

\email{rodriguez.puebla@gmail.com}

\begin{abstract} 
By means of a statistical approach that combines different semi-empirical methods of galaxy-halo connection,
we derive the stellar-to-halo mass relations, \shmr, of local blue and red central galaxies separately. We also 
constrain the fraction of halos hosting blue/red central galaxies and the occupation statistics of blue and red satellites 
as a function of halo mass, \mh. For the observational input, we use the blue and red central/satellite
galaxy stellar mass functions and two-point correlation functions in the stellar mass range of $9<$log(\ms/\msun)$<12$. 
We find that: (1) the \shmr\ of central galaxies is segregated by color, with blue centrals having
a \shmr\ above the one of red centrals; at log(\mh/\msun)$\sim 12$, the \ms-to-\mh\ ratio of the blue centrals is 
$\approx 0.05$, which is $\sim1.7$ times larger than the value of red centrals. 
(2) The constrained scatters around the $\shmr$s of red and blue centrals are $\approx0.14$ and $\approx0.11$ dex, respectively.    
 The scatter of the average \shmr\ of all central galaxies changes from $\sim0.20$ dex to $\sim0.14$ dex in the 
 $11.3<$log(\mh/\msun)$<15$ range. 
(3) The fraction of halos hosting 
blue centrals at $\mh=10^{11}$ \msun\ is 87\%, but at $2\times 10^{12}$ \msun\ decays to $\sim20\%$,
approaching to a few per cents at higher masses. The characteristic mass at which this fraction is the same
for blue and red galaxies is $\mh\approx 7\times10^{11}$ \msun.
Our results suggest that the \shmr\ of central galaxies at large masses is shaped by mass quenching.   
At low masses, processes that delay star formation without invoking too strong supernova-driven outflows
could explain the high \ms-to-\mh\ ratios of blue centrals as compared to those of the scarce red centrals.
\end{abstract}


\section{Introduction}

The current paradigm of galaxy formation and evolution has its theoretical background in the
$\Lambda$ cold dark matter (\lcdm) cosmological model. In this paradigm, 
the backbone of galaxy formation are the gravitationally bound CDM structures (halos) in 
the cosmic web. The statistical
properties and mass assembling of CDM halos have been calculated 
with great detail, mainly by means of large N-body cosmological simulations 
\citep[for a recent review, see][]{Knebe+2013}. Of particular relevance is the 
halo mass function (HMF), the number of halos of a given mass per unit of comoving volume,
which can be divided into halos not contained inside larger ones (distinct) and subhalos.
In the understanding that halos and subhalos are populated respectively by central 
and satellite galaxies,
the galaxy-(sub)halo connection can be stablished at a {\it statistical level} by using observed
galaxy distributions such as the galaxy stellar mass function 
(\gsmf), the two-point correlation function, and the satellite conditional stellar mass functions.
The resulting semi-empirical galaxy-(sub)halo connection provides a powerful tool to constrain
galaxy evolution models as well as the properties of galaxies as a function of scale and 
environment \citep[][]{Mo+2010}. 

In recent years, several statistical approaches have emerged for connecting galaxies to
their CDM halos. Among the simplest ones is the
so-called (sub)halo abundance matching technique (\amt). The \amt\ consists in
assigning by rank a galaxy stellar mass, \ms, (or luminosity) to a host dark matter halo of mass \mh\ 
by matching their corresponding cumulative number densities. 
\citep[e.g.,][]{Kravtsov+2004,ValeOstriker2004,ValeOstriker2008,Conroy+2006,Shankar+2006,
Behroozi+2010,Guo+2010,RDA12,Papastergis+2012,Hearin+2013,Behroozi+2013}. 
As a result from this matching, one obtains the total stellar-to-halo mass relation (\shmr). 
Usually the \shmr\ is assumed to be identical both for central and for satellite galaxies. However, 
recent studies \citep{Neistein+2011,Yang+2012,RDA12,Reddick+2013} have questioned this 
assumption, which has intrinsic issues
for satellite galaxies and implicitly assumes that their $\shmr$ does not evolve as a function
of redshift and they have the same evolution trajectories as subhalos 
(e.g., no orphan satellite galaxies, etc.). 
In addition to the \amt, there are other
semi-empirical approaches for constraining the distribution of central and satellite galaxies inside the halos,
such as the halo occupation distribution (HOD) model \citep[e.g.,][]{Jing+1998,Berlind+2002, Cooray+2002,
Zehavi+2005,Abbas+2006,Foucaud+2010,Zehavi+2011,Watson+2011,Wake+2012,Leauthaud+2012},
and the closely related conditional stellar mass (or luminosity) function model 
\citep[][hereafter \citetalias{Yang+2012}]{Yang+2003,vandenBosch+2003,Cooray+2006,Yang+2007,vandenBosch+2007,Yang+2009b,Yang+2012}. 
These approaches use the observed 2PCF and/or galaxy group catalogs for constraining the central/satellite galaxy distributions, respectively.
By combining thus constrained survived satellite population together with accreted ones predicted using the
halo merger histories, one can model the evolution of these galaxies satellite galaxies and their contribution to the
evolution of central galaxies  \citep[e.g.,][]{Yang+2009a, Yang+2012,Yang+2013}

By combining the above mentioned semi-empirical approaches, \citet[][hereafter RAD13]{RAD13}
were able to derive separately the \shmr\ of local central galaxies--distinct halos and satellite 
galaxies--subhalos as well as several occupation distributions \citep[for closely related
works see also][]{Yang+2009a,Neistein+2011,Reddick+2013,Watson+2013}. Actually,
the total, central, and satellite \shmr s have intrinsic scatters related to 
the stochastic halo assembly and the complex processes of galaxy evolution. In most of the
previous works, the intrinsic scatter around the median \shmr\ has been assumed as random
as well as constant as a function of halo mass. Previous works have measured or 
constrained the intrinsic scatter around the \shmr\ 
finding typically that this is small, $\sim 0.15-0.20$ dex in log\ms \citep[ c.f.][\citetalias{RAD13}]
{Mandelbaum+2006,Yang+2008,More+2011,Skibba+2011,Leauthaud+2012,Reddick+2013,
Kravtsov+2014}.

A natural next step in understanding the link between galaxies and halos is to explore 
what  galaxy properties are related to the shape and scatter of the \shmr. The fact that for 
a given \mh, there are galaxies more or less massive than the mean \ms\ corresponding to this
\mh, certainly tells us something about the galaxy evolution process, in particular if these deviations 
correlate with a given galaxy property. In the era of big galaxy surveys, galaxy color
is one of the most immediate observational properties reported in these surveys. It is well known 
that the color distribution of galaxies is bimodal  \citep[c.f.][]{Baldry+2004,
Brinchmann+2004,Weinmann+2006,vandenBosch+2008}.
This color bimodality strongly correlates with mass and in less degree with environment 
\citep[][and more references therein; see also \citealp{Peng+2010}]{Blanton+2009}. 
The color is a fundamental property of galaxies related mainly to their star formation (SF) history,
and it correlates in more or less degree with other galaxy properties such as the specific
star formation (SF) rate and morphology.

Here, we pose the question {\it whether the \shmr\ of local blue and red central galaxies are similar or not},
and therefore, whether the scatter around the total \shmr\ of central galaxies is segregated
by color. In order to tackle this question as general as possible, the assumption that the distribution
of \ms\ for a given \mh\ is given by a unique random (lognormal) function should be relaxed; instead we 
will consider that blue and red galaxies have their own distributions (intrinsic scatters). If both
distributions, after being constrained with observations, are statistically similar, then galaxy
color is not the responsible for shaping the intrinsic scatter of the \shmr. 

Researchers have attempted to constrain the 
galaxy-halo connection for local and high-redshift galaxies separated by color or morphology 
using direct methods, namely the
galaxy-galaxy weak lensing \citep{Mandelbaum+2006, vanUitert+2011,Velander+2014,Hudson+2014} 
and the satellite kinematics \citep{Conroy+2007,More+2011,Wojtak+2013}
In order to attain the necessary signal-to-noise ratio,
the current method requires stacking the data from large surveys, and 
even then uncertainties are yet large. Despite the large 
uncertainty and limited mass range, the obtained results suggest that {\it the \shmr\ of 
blue and red (late- and early-type) galaxies could be different}.  Additionally, in a combined
analysis of galaxy clustering and galaxy-galaxy weak lensing, \citet{Tinker+2013} also 
concluded that the \shmr\ of passive and active galaxies are different in the COSMOS field at 
$z=[0.2,1.0]$ \citep[see also][]{Hartley+2013}.

By using a semi-empirical model that generalizes the one presented in \citetalias{RAD13}, here
we determine statistically the local \shmr\ of blue and red central galaxies separately, in the
mass range of $\ms\approx 10^9-10^{12}$ \msun, as well as their corresponding blue and red
satellite populations. In addition, we constrain separately the 
scatters around the $\shmr$s of blue and red centrals, and the scatter 
around the (bimodal) distribution of the (average) \shmr\ of all central galaxies. 
The semi-empirical study of the \shmr\ of {\it central} blue/red galaxies will allow us   
to understand what mechanisms carved the \shmr\ and will shed light into the 
relevant galaxy evolutionary processes as a function of scale and environment. Additionally,
an important aspect of the \shmr\ of blue and red central galaxies is  whether they are consistent with  
observabed galaxy correlations as the Tully-Fisher and Faber-Jackson relations. 
In the present paper we present our semi-empirical model in detail and the main results
regarding the \shmr\ of central blue and red galaxies. In future works, we will use
the model results for exploring the mentioned above aspects. 
In addition, we will also explore the properties of halos hosting blue and red central galaxies. 

The plan for this paper is as follows. In Section \ref{model}, we describe our
semi-empirical approach, as well as the required input, key assumptions, 
and the statistical procedure for constraining the model parameters. 
In Section \ref{data}, we describe the observational data to be used. 
Our main results are presented in Section \ref{results}. In Section \ref{Discussion}, we discuss
on the robustness and the interpretation of our results. In the same Section, we also compare
the results obtained here with previous studies.
Finally, a summary and the conclusion are presented in Section \ref{conclusions}.  

Unless otherwise stated, all of our calculations are based on a flat $\Lambda$CDM cosmology with 
$\Omega_{\Lambda}=0.73$, $h=0.7$ and $\sigma_8=0.84$.

\section{The semi-empirical model}
\label{model}

In this Section we describe the semi-empirical model developed for connecting 
blue and red central galaxies to their host dark matter halos, and 
for obtaining the occupational statistics of blue and red satellite galaxies. 
The model allows us to relate the central and satellite 
\gsmf s and the projected two-point correlation functions (\pcf s), as well as their 
decompositions into blue and red galaxies, to the theoretical \lcdm\ halo mass function. 
By means of this model, from the observed total, central and satellite \gsmf s and the 
projected \pcf s, in all the cases decomposed into blue and red populations,
we can constrain: the stellar-to-halo mass relations, \shmr s, of blue, red and all (average) 
central galaxies, the fraction of halos hosting blue and red central galaxies, 
and the satellite blue/red conditional stellar mass functions (\cmf s) as a function 
of host halo or central galaxy stellar mass. 
The statistical model presented here combines the \amt, \hod\ model and 
\cmf\ formalism as presented in \citetalias{RAD13}. 
In order to include separately populations of blue and red 
galaxies, one requires some additional ingredients described as follows:

\begin{itemize}
\item For connecting blue and red central galaxies to their host dark matter halos, we introduce 
the conditional probability distribution functions that a distinct halo of mass \mh\ hosts either a
blue or red central galaxy in the stellar mass bin $\ms\pm d\ms/2$, 
denoted by \Pbc\ and \Prc, respectively. As a result, these 
distributions contain information about the \shmr\ (mean and scatter) of blue and 
red central galaxies, \mbc(\mh) and \mrc(\mh), respectively. 

\item In order to derive \mbc(\mh) and \mrc(\mh), the fraction of halos hosting blue and red 
central galaxies should be known. Motivated by observational results, we 
introduce a parametric function for these fractions that will be constrained with
the observational input.

\item To model the occupational numbers of blue and red satellites, we
use observationally-motivated parametric functions for the blue and red 
satellite \cmf s. 

\end{itemize}

\begin{figure*}
\vspace*{0pt}
\hspace*{50pt}
\includegraphics[height=6.1in,width=3in, angle=-90]{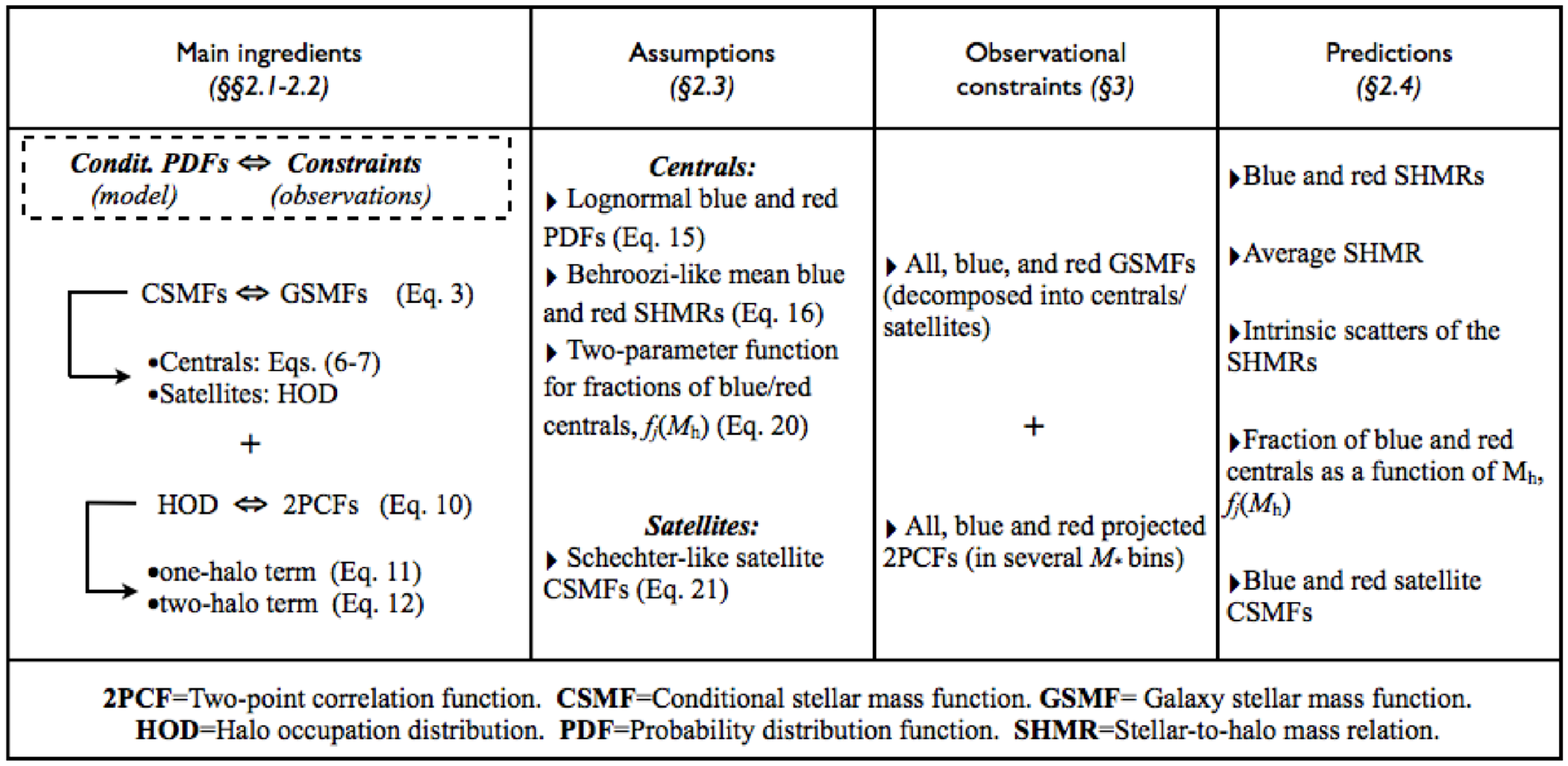}
\caption{Summary of our model, its main assumptions, observational data required for
constraining it, and the main predictions from this model. The relevant (sub)sections
and equations are indicated.
 }
\label{scheme}
\end{figure*}

In Fig. \ref{scheme}, we present a schematic table that summarizes
the main idea behind our model. We also indicate the kind of observational data   
we use for constraining the model parameters as well as our model predictions. 
Note that relevant (sub)sections and equations are also indicated.
In the following we describe our model in more detail as well as 
the observational data employed to constrain the model parameters. 

Those readers interested only on the results obtained with our model and their implications 
may prefer to skip to Section \ref{results}. 

\subsection{Modeling the Galaxy Stellar Mass Function}
\label{model_gsmf}

One can express the total \gsmf\ by separating the population of
galaxies into central and satellites:
\begin{equation}
\phig(\ms)=\phigc(\ms)+\phigs(\ms),
\end{equation}
each of which can be subdivided into blue and red galaxy populations, i.e.,
\begin{equation}
\phig(\ms)=[\phigbc(\ms)+\phigrc(\ms)]+[\phigbs(\ms)+\phigrs(\ms)].
\end{equation}

By defining the \cmf, \Phiij, as the mean number of `$i$ type' ($i=$central or
satellite) galaxies of a `$j$ color' ($j=$blue or red) at the mass
bin $\ms\pm d\ms/2$, one can write each component of the {\gsmf}s in the 
following form:
\begin{equation}
\phigsij(\ms)=\int\Phiij\phih(\mh)d\mh,
\label{phij}
\end{equation}
where \phih\ is the {\it distinct} halo mass function.
Thus, the mean cumulative number density galaxies of type `$i$' and color `$j$'  
can be written as
\begin{equation}
\ngij(>\ms)=\int\Nsij\phih(\mh)d\mh,
\label{nij}
\end{equation}
where,
\begin{equation}
\Nsij=\int_{M_*}^\infty\Phiijint dM_*',
\label{Nij}
\end{equation}
is the mean cumulative number of galaxies of the type `$i$' and color `$j$'  
with stellar masses greater than $\ms$ residing in a halo of mass \mh.
Observe that once the $\cmf$s $\Phiij$ are given, Eqs. (\ref{phij}-\ref{Nij})
are totally defined. Therefore, {\it the key ingredients in our model are the conditional
mass functions} $\Phiij$.

\subsubsection{Central Galaxies}
\label{model_centrals}

In the context of the \amt, the connection between the total central \gsmf, \phigc(\ms), 
and the distinct halo mass function, \phih(\mh), arises naturally
by assuming a probability distribution function, denoted
by $\Pcen$, that a distinct halo of mass \mh\ hosts a central galaxy 
in the stellar mass bin $\ms\pm d\ms/2$ (see Introduction for references).
As a result of this connection, the mean \shmr\ of central galaxies, \ms(\mh), 
can be constrained.
In the case that the \gsmf\ is divided into different populations, 
the above idea can be extended 
in order to connect the different galaxy populations to their host dark matter halos. 
For blue and red galaxies, one can introduce the conditional probability 
distribution functions \Pbc\ and \Prc\ to establish the statistical connection 
between the ``blue", \phibh, and ``red", \phirh, distinct halo mass functions and the $\gsmf$s of 
blue and red centrals, \phigbc\ and \phigrc, respectively. As above, 
the mean relations \mbc(\mh) and \mrc(\mh) are the result of this connection. 

In terms of the \cmf\ formalism, one can specify the central \cmf\ as the sum of 
the blue and a red components, 
\begin{equation}
\Phicen=\Phicenb+\Phicenr,
\label{csmf_cen}
\end{equation}
where the \cmf\ of blue and red central galaxies are given by
\begin{equation}
\Phicj=\fj(\mh)\times\Pcj.
\label{csmf_j}
\end{equation}
As above, the subscript $j$ refers either to red (r) or blue (b) central galaxies, 
and $\fj(\mh)$ {\it is the fraction of halos hosting central galaxies of color $j$.} 
Notice that for all central galaxies, the \cmf\ is simply given by $\Phicen=\Pcen$. 

By inserting Eq. (\ref{csmf_j}) into Eq. (\ref{csmf_cen}) one can obtain  
the relation between the probability 
distribution functions \Pcen,  \Pbc\ and \Prc,
\begin{eqnarray}
\Pcen=\Pbc\fb(\mh)+&  &\nonumber \\
\Prc\fr(\mh).
\label{pcen}
\end{eqnarray}

In Section \ref{central-galaxies}, we discuss the parametric functional forms for 
each $\Pcj$. In addition, based on the results of galaxy groups we also motivate the functional form for
of each $\fj(\mh)$.

\subsubsection{Satellite Galaxies}
\label{model_sats}

As mentioned above, in our model the total distribution of satellite galaxies 
will be characterized by means of the satellite \cmf, \Phisat. 
Similarly, the distribution of blue and red satellite galaxies will be characterized
by means of  \Phisatj, where the subscript $j$ stands for either blue 
(b) or red (r) galaxies. 
As we will discuss in Section \ref{sat-gals}, the  parametric functional forms employed
for each \Phisatj\ are motivated by previous empirical results of galaxy groups. 
From these definitions, it follows that at a fixed \mh, 
the mean fractions of blue and red satellites
as a function of \ms\ are: 
\begin{equation}
\fjsat=\Phisatj/\Phisat,
\label{fbsat}
\end{equation}
where by definition $\frsat+\fbsat=1$.

\subsection{The correlation function in the HOD model}
\label{model_clustering}

Once the link between blue and red  central galaxies to their host halos and 
the satellite \cmf s have been specified, we can proceed to compute the 
spatial clustering of galaxies as a function of stellar mass and color by using 
the HOD model. This connection is introduced in order {\it to use the 
observed \pcf s as constraints to the model parameters}.

In the HOD model  (see Introduction for references), the real space \pcf\ is computed 
by decomposing it into two parts, the one-halo term at small scales, 
and the two-halo term at large scales. Here, we 
model the real space \pcf\ for `$j$'-galaxies, i.e., either for all,
blue or red galaxies as,
\begin{equation}
1+\xij=[1+\xiggj]+[1+\xiggggj].
\end{equation}
The one-halo
term describes the number of all possible pairs coming from galaxies in same halos, 
while the two halo term describes the same but in separate halos. 
Specifically, in the HOD model
context the one-halo term is given by,
\begin{eqnarray}
\frac{\Nij}{2}\lnfw=\Ncjj\Nsatjj\lnfwcs&  &\nonumber \\
+\frac{1}{2}\Nssij\lnfwss,
\label{1_halo}
\end{eqnarray}
where the \ms\ and \mh\ dependences in the mean cumulative numbers 
defined above (see Eq. \ref{Nij}) were omitted for simplicity. The term \lnfw\
is the spatial distribution of the galaxies within the dark matter halo. 
In Eq. (\ref{1_halo}), we have assumed 
that central-satellite pairs follow a pair distribution function 
$\lambda_{c,s}(r)dr=4\pi\tilde{\rho}_{\rm NFW}(\mh,r) r^2dr$, where
$\tilde{\rho}_{\rm NFW}(\mh,r)$ is the normalized NFW halo density profile.   
The satellite-satellite pair distribution, $\lambda_{s,s}(r)dr$, is then the
normalized density profile convolved with itself, that is,
$\lambda_{s,s}(r)dr=4\pi{\lambda}_{\rm NFW}(\mh,r) r^2dr$,
where ${\lambda}_{\rm NFW}$ is the NFW profile convolved with itself. 
An analytic expression for ${\lambda}_{\rm NFW}(\mh,r)$ is given by 
\citet{Sheth+2001a}. Both $\tilde{\rho}_{\rm NFW}$ and ${\lambda}_{\rm NFW}$ 
depend on the halo concentration parameter, $c_{\rm NFW}$. N-body numerical 
simulations show that this parameter anti-correlates with mass, 
$c_{\rm NFW}=y_0 - y_1\times$log\mh, though with a large scatter. Note that we have
assumed that the occupational number of satellite galaxies follows a Poisson distribution,
i.e., $\Nss=\Nssij$. The above is based on the results of high-resolution $N$-body
simulations \citep[e..g,][]{Kravtsov+2004} and hydrodynamic simulations of galaxy formation 
\citep[e.g.,][]{Zheng+2005}. 

On large scales, we model the two halo-term as
\begin{equation}
\xiggggj=b^2_{g,j}\zeta^2(r)\xi_{mm}(r),
\end{equation}
where $\xi_{mm}(r)$ is the nonlinear matter correlation function \citep{Smith+2003},
$\zeta(r)$ is the scale dependence of dark matter halo bias \citep[][see their Eq. B7]{Tinker+2005},
and
\begin{equation}
b_{g,j}=\frac{1}{n_g}\int b(\mh)\Navej\theta(r;R_{vir}(\mh)) \phi(\mh)d\mh
\label{bias}
\end{equation}
is the galaxy bias. In the above Eq. (\ref{bias}), $b(\mh)$ is the halo bias function given by
\citet{Tinker+2010}. 
The term $\theta(r;R_{vir}(\mh))$ is the Heaviside function and has been 
introduced to take into account 
that two galaxy pairs cannot be within the same halo. \citet{Wang+2004} showed that
the above method describes accurately well the correlation function 
(see also \citetalias{Yang+2012}). 
Analogously to \citet{Leauthaud+2012}, 
we have modified the original \citet{Wang+2004} method to match our definition
of halo mass functions and bias relation. This fitted relation have been obtained based on 
spherical-overdensity halo finding algorithms, where
halos are allowed to overlap as long as their centers are not contained 
inside the virial radius of
a larger halo; for details see \citet[][]{Tinker+2010}. 

Observations of galaxy clustering are usually
characterized by using the galaxy projected 
correlation function, $\omega_{\rm p}(r_{\rm p})$. 
In our model, we relate $\omega_{\rm p}(r_{\rm p})$ 
to the real-space correlation function, $\xi_{gg}(r)$, by the integration
over the line of sight:
\begin{equation}
\omega_{{\rm p},j}(r_{\rm p})=2\int_0^{\pi_{\rm max}}\xi_{gg,j}(\sqrt{r^2_{\rm p}+r_{\pi}^2})dr_{\pi}.
\end{equation}
For consistency with the observed $\omega_{\rm p}(r_{\rm p})$ we set 
$\pi_{\rm max}=45$ Mpc $h^{-1}$ .

\subsection{Model assumptions}
\label{inputs}

In order to constrain the model, some assumptions for the different distributions should be made. 
In this subsection we describe these assumptions in detail.

\subsubsection{Central galaxies}
\label{central-galaxies}

As we have noted in subsection \ref{model_gsmf}, our model for central galaxies is completely 
specified once the \cmf\ of blue and red central galaxies are defined, 
see Eqs. (\ref{csmf_cen}-\ref{pcen}). Here, both \Pbc\ and \Prc\ 
are modeled as lognormal distributions:
\begin{eqnarray}
 	 \Pcj d\ms=\frac{\log e}{\sqrt{2\pi{\sigma_{j}}^2}}\times& & \nonumber \\
 		 \exp\left[{-\frac{\log^2(\ms/M_{*,j}(\mh))}{2\sigma_{j}^2}}\right]\frac{d\ms}{\ms},
  \label{cmfc}
\end{eqnarray}
 where $M_{*,j}(\mh)$ is either the mean \shmr\ of blue, \mbc(\mh), or red, \mrc(\mh), central
 galaxies, and the standard deviations \sigmaj's are defined here 
 as the corresponding scatters around the mean relations. 
 We assume that the \sigmaj's are independent of \mh.  
Both $\sigma_{b}$ and $\sigma_{r}$ are considered as additional parameters 
to be fitted separately in the model. 

The scatters \sigmaj\ are composed by an
{\it intrinsic component} and by a {\it measurement error component}.
The measurement error components, $\sigma_{j}^{\rm e}$, are dominated mainly
due to errors in individual galaxy stellar mass estimates and redshift estimates.
\citep[see e.g.,][]{Behroozi+2010,Leauthaud+2012}. Additionally, their value may depend 
on galaxy color \citep{Kauffmann+2003}. Thus, if the measurement error components 
are known, it is then possible to constrain the intrinsic components in our model, 
$\sigma_{j}^{\rm i}$. However, given the poor information for the real values of
$\sigma_{j}^{\rm e}$, in our analysis we opt for constraining total scatters, \sigmaj, as 
upper limits to the intrinsic components. Nevertheless, in Section \ref{scatter} 
we estimate conservative values by assuming that both components are
independent, 
\begin{equation}
\sigma_{j}^2 = (\sigma_{j}^{\rm i})^2 + (\sigma_{j}^{\rm e})^2.
\label{sigmasj}
 \end{equation}
 and a constant value for $\sigma_{j}^{\rm e}$ reported in \citet{Kauffmann+2003}.

 In order to describe the mean \shmr\ of blue and red central galaxies, we adopt the parametrization 
 proposed in \citet{Behroozi+2013},
\begin{eqnarray}
	  \log M_{*,j}=\log(\epsilon_{c,j} M_{1,j})+g_c(x)-g_c(0),
 		 \label{msmh}
\end{eqnarray}
where
\begin{equation}
	 g_c(x)=\delta_{c,j}\frac{(\log(1+e^x))^{\gamma_{c,j}}}{1+e^{10^{-x}}}-\log(10^{\alpha_{c,j} x}+1).
		 \label{msmh1}
\end{equation}
and $x=\log(\mh/M_{1,j})$.
This function behaves as a power law with slope $\alpha_{c,j}$ at masses much
smaller than $M_{1,j}$, and as a sub-power law with slope $\gamma_{c,j}$ at
large masses. 
A simpler function, with less parameters could be used (e.g., \citealp{Yang+2008}),
however, as shown in \citet{Behroozi+2013}, the function as given by Eq. (\ref{msmh1}) is
 necessary in order to map accurately the \hmf\ into the observed \gsmf s, which are more 
 complex than a singular Schechter function (see \S\S \ref{gsmfs-obs} below).

Deviating from previous studies, in our model {\it the probability distribution 
for all central galaxies, \Pcen}(Eq. \ref{pcen}), {\it is predicted} rather than being an
assumed prior function \citep[see also][]{More+2011}. Typically, this distribution is assumed
as a lognormal function with a fixed width. In our model, by means of Eq. (\ref{pcen}), the
mean \shmr\ of all central galaxies, \ms(\mh), is given by the weighted sum 
of the mean blue, \mbc(\mh), and red, \mrc(\mh), central $\shmr$s,
\begin{eqnarray}
	\langle\log\ms(\mh)\rangle=\fb(\mh)\langle\log\mbc(\mh)\rangle& & \nonumber \\
		+\fr(\mh)\langle\log\mrc(\mh)\rangle.
		\label{mean_shmr}
\end{eqnarray}
Observe that this equation relates the mass relation commonly obtained through the HOD
model and the \cmf\ formalism with the mass relations of blue and red centrals. 
We also compute the intrinsic scatter around the average relation as:
\begin{equation}
	\sigma_A(\mh)=\left(\int P_{c,t}(\mathcal{M_*}|\mh)\mu^2d\mathcal{M_*}\right)^{1/2},
		\label{sgima_msmh}
\end{equation}
where $\mu=\log\mathcal{M_*}-\langle\log\ms(\mh)\rangle$. Note that we are {\it not
assuming} that the scatter around the mean \shmr\ of all central galaxies 
is constant and lognormally distributed. Because the value $\sigma_A(\mh)$ is directly
related to the \sigmaj's, it is also a combination of the intrinsic and
error measurement components (see Eq. \ref{sigmasj}). 

To fully characterize the $\cmf$s of blue and red central galaxies,
we need to propose a model for the fraction of halos hosting blue/red central galaxies, $\fj(\mh)$
(see Eq. \ref{csmf_j}).
As noted by previous authors 
\citep[c.f.][]{Hopkins+2008,TinkerWetzel2010,Rodriguez-Puebla+2011,Tinker+2013}, assuming a 
specific function of the quenched (red) fraction makes an implicit choice 
of the mechanisms that prevent central galaxies to be actively star-forming.
For example, in \citet{Rodriguez-Puebla+2011}, the fraction of halos able to host blue centrals was obtained
by excluding from the \lcdm\ halo mass function (1) those halos that suffer a major merger at 
$z<0.8$, and (2) those that follow the observed rich group/cluster mass function (blue galaxies
are not found in the center of rich groups/clusters).

In a recent analysis of the \citet{Yang+2007} galaxy group catalog,  
\citet{Woo+2013} studied the fraction of quenched central galaxies as a 
function of both stellar and halo mass. From their analysis, the authors 
concluded that the fraction of quenched central galaxies correlates stronger with
\mh\ than with \ms. Furthermore, \citet{Woo+2013}  concluded that the 
phenomenological results presented in \citet{Peng+2012} for central galaxies
are still valid by substituting \mh\ for \ms\ in the \citet{Peng+2012} model. 
In other words, central galaxies are 
on average quenched once their host dark matter halo reaches a characteristic mass. 
According to the phenomenological results of \citet{Peng+2012} and \citet{Woo+2013},
the fraction of halos hosting red (or blue) centrals can be described as 
$\fr(\mh)=1/(1+[\mhchar/\mh])$, where \mhchar\ is a characteristic mass above which 
halos are mostly occupied by red central galaxies; for $\mh>\mhchar$,                 
$\fr>0.5$ (or $\fb=1-\fr<0.5$).        
Given the central role that this function plays in our model, we have slightly generalized it as follows:
\begin{equation}
	\fr(\mh)=\frac{1}{b+(\mhchar/\mh)},       
		\label{fb-Mhalo}
\end{equation}
where \mhchar\ and $b$ are free parameters to be constrained.
For practical purposes we redefine $\mhchar=\beta\times10^{12}\msun$,
where $\beta$ is the free parameter to be constrained.

For the distinct halo mass function, we use the fit to large N-body cosmological simulations 
presented in \citet{Tinker+2008a}. 
Here we define halo masses at the virial radius, i.e.\ the radius where 
the spherical overdensity is $\Delta_{\rm vir}$ 
times the mean matter density, with $\Delta_{\rm vir}=(18\pi^2+82x-39x^2)/\Omega(z)$, and
$\Omega(z)=\rho_m(z)/\rho_{\rm crit}$ and $x=\Omega(z)-1$. 

Finally, for the relation of the halo concentration parameter $c_{\rm NFW}$ with
mass, we use the fit to N-body cosmological simulations by \citet[][]{Munoz+2011}.

\subsubsection{Satellite galaxies}
\label{sat-gals}

The parametric functions for describing the satellite \cmf s, \Phisatj,  to be used here   
are given through the average satellite cumulative probabilities:
\begin{equation}
	\Phisatj=\frac{\partial}{\partial\ms}\Nsatj,
		\label{Phisatb}
\end{equation}
where
\begin{equation}
	\Nsatj=\Ncen\int_{M_*}^\infty S_j(M_*'|\mh)dM_*',
		\label{cum_blue}
\end{equation} 
and
\begin{equation}
	S_j(\ms|\mh)=\phi^*_{j}X^{\alpha_{s,j}}e^{-X}dX,
		\label{Sb}
\end{equation}
and $X=\ms/\msatj$. As before, the subscript `$j$' refers either to
red (r) or blue (b) galaxies. Note that the first factor in Eq. (\ref{cum_blue}) 
(the average cumulative probability of having a central galaxy larger than \ms\  
in a halo of mass \mh) imposes
the restriction that there are not galaxy groups containing only satellite galaxies
and that, on average, the central galaxy is the most massive galaxy in the group. 
In the above equation we assume that the faint-end slopes $\alpha_{s,j}$ are
independent of halo mass, while we the normalizations factors $\phi^*_{j}$ and the
characteristic masses $\msatj$ change as a function of \mh\ as follows,
\begin{equation}
	\log\phi^*_{b}(\mh)=\phi_{0,j}+\phi_{1,j}\times\log\left(\frac{\mh}{10^{12}\msun}\right),
\end{equation}
and
\begin{equation}
	\log\msatj = c_0 + c_j\times \log\left(\frac{\mh}{10^{12}\msun}\right).
		\label{Msatj}
\end{equation}
respectively. Note that the normalization in the last Eq., $c_0$, is the same for blue and red satellites.  

The parametrization presented above is  partially motivated by the phenomenological 
model discussed in \citet{Peng+2012}. These authors argue that the shape of the distribution of 
blue/star-forming satellite galaxies is always a Schechter function with a characteristic
mass \msatb\ and a faint-end slope $\alpha_{s,b}$. In the case of the distribution of red/quenched
galaxies, they argued that it is described by a double Schechter with a characteristic mass similar 
to that of blue satellites and with slopes $\alpha_{s_1,r}=1+\alpha_{s,b}$ and $\alpha_{s_2,r}=\alpha_{s,b}$. 
 We have experimented with the cases of a simple and a double Schechter function
and, in the light of the observations we use to constrain the model 
(Section \ref{procedure}), there is no a statistical improvement in the fittings
from one to the other case. We have also checked that the \cmf s of red 
satellites from the \citetalias{Yang+2012} galaxy group catalog can be fitted both with a double
or a simple Schechter function with $\alpha_{s,r}$. Therefore, 
we parametrize the red satellite \cmf s with a simple Schechter function. 
In order to tackle this question as 
general as possible, we have assumed that  $\alpha_{s,b}$ and  $\alpha_{s,r}$ are
two different parameters.

\subsection{The procedure}  
\label{procedure}

\subsubsection{Parameters in the model}

We now summarize the set of free parameters defined in our phenomenologically
motivated model:
$\vec{p}=(\vec{p}_b,\vec{p}_r,\vec{p}_\sigma,\vec{p}_{f_b}, 
\vec{p}_{\rm CSMF})$.
Five parameters are to model the \shmr\ of blue central galaxies, 
$\vec{p}_b=(\epsilon_b,M_{1,b},\alpha_b,\delta_b,\gamma_b)$, and five to model the 
\shmr\ of red central galaxies, $\vec{p}_r=(\epsilon_r,M_{1,r},\alpha_r,\delta_r,\gamma_r)$ 
(Eqs. \ref{msmh} and \ref{msmh1}); two more parameters are to constrain the (assumed lognormal)
scatter around each \shmr, $\vec{p}_\sigma=(\sigma_b,\sigma_r)$. 
Two parameters correspond to 
the function used for constraining the fraction of halos hosting red central galaxies, 
$\vec{p}_{f_r}=(\beta, b)$, see Eq. (\ref{fb-Mhalo}). 
Finally, nine  parameters are to constrain the blue and red satellite \cmf s, 
$\vec{p}_{\rm CSMF}=(\alpha_{s,b},,\phi_{0,b},\phi_{1,b},c_0, c_b, c_r,\alpha_{s,r},\phi_{0,r},\phi_{1,r})$, 
see Eq. (\ref{Sb}).

\subsubsection{Fitting procedure}

In order to constrain the free parameters in our semi-empirical model,
we combine several observational data sets. These data sets are the 
\gsmf s and its division into central and satellites and 
the \pcf s in different stellar mass bins for all, blue and red galaxies.
In order to sample the best-fit parameters that maximize the likelihood 
function $L\propto e^{-\chi^2/2}$ we use the 
Markov Chain Monte Carlo (MCMC) method.
The details for the full procedure can be found in \citetalias{RAD13}.

We compute the total $\chi^2$ as,
\begin{equation}
	\chi^2=\chi^2_{\rm GSMF}+\chi^2_{\rm 2PC} 
\end{equation}
where for the $\gsmf$s we define,
\begin{equation}
	\chi^2_{\rm GSMF}=\sum_{i,j}\chi^2_{\phi_{i,j}},
\end{equation}
the sum over $i$ refers to the type (all, centrals, and satellites) while
the sum over $j$ refers to color (blue and red). 
For the correlation functions,
\begin{equation}
	\chi^2_{\rm 2PC}=\sum_k\left(\chi^2_{\omega_k(r_p)_t}+\chi^2_{\omega_k(r_p)_b}+
		\chi^2_{\omega_k(r_p)_r}\right),
\end{equation}
where the subscripts `t', `b' and `r' refer to all (total), blue and red galaxies, respectively.  
The sum over $k$ refers to summation over different stellar mass bins.           
The fittings are made to the data points (with their error bars) for each \gsmf\ and \pcf.          

Once the model parameters are constrained, the model {\it predicts}
the following relations and quantities: 
\begin{enumerate}
\item The mean $\shmr$s of blue and red central galaxies as well as of 
the average \shmr\ of all central galaxies. The latter is what is commonly constrained in 
the literature through the \amt.    
\item The scatter around the blue, red and average \shmr s of central galaxies. 
Recall that we assume lognormal distributions with constant widths (i.e., scatters independent of \mh)
for blue and red centrals. In contrast, the distribution for all central galaxies will be predicted
according to Eq. (\ref{pcen}). Similarly, the  scatter around the mean \shmr\ is predicted
according to Eq. (\ref{sgima_msmh}).  Note that these scatters are upper limits to 
the intrinsic scatters. These  can be estimated if the values for the measurement error components are 
known (see Eq. \ref{sigmasj}).
\item The fraction of distinct halos hosting red (or blue) central galaxies as a function of \mh, \fr(\mh).   
\item The blue and red satellite \cmf s (as a function of \mh\ or central \ms),  and therefore
the fractions of blue (or red) satellites as a function of \ms\ for a given host halo mass, \fbsat.
\end{enumerate}

\subsection{Comparison to other models}  
\label{ctom}

The model described in this Section for constraining 
$\shmr$s separately for blue and red central galaxies  
is partially related to some previous models discussed in the literature. 
Following, we outline the main differences between our and
previous models. 

As discussed in the Introduction, for connecting  central galaxies to their host
dark matter halos, previous models assumed that the probability distribution of \ms\ for a given 
\mh, is given by an unimodal (lognormal) distribution function (see, e.g., \citetalias{Yang+2012}; 
\citealp{Leauthaud+2012})
In addition to that, in previous models, the scatter around the mean \shmr\ 
was assumed to be constant with halo mass. In our approach, we relax these assumptions 
by assuming separate probability (lognormal) distribution functions for blue and red
centrals in such a way that the average (density-weighted) distribution function 
can be now bimodal and formally may depend on \mh. 
This is similar to the model employed in \citet{More+2011}
for obtaining \shmr s for blue and red central galaxies based on the analysis of the 
satellite kinematics from a local sample in the SSDS, 
and that of \citet{Tinker+2013} based on a combined analysis
of galaxy clustering and weak lensing for active and passive galaxies in the COSMOS field. 
It is worth to remark that {\it if both distributions after being constrained with observations
are found to be statistically similar, then the galaxy color is not responsible for shaping 
the intrinsic scatter of the \shmr.} In the opposite case, the result is evidence that the \shmr\ 
is segregated by color.

In essence, we find that our model is closer to the one developed in \citet{Tinker+2013}. The main 
difference resides on the fact that these authors characterize the galaxy distribution
using the HOD model \citep[see also][]{Leauthaud+2012}, while in our analysis this characterization 
is based on the \cmf\ formalism 
\citep[see e.g,][]{Yang+2003,vandenBosch+2003,Cooray+2006,Yang+2012}. Note, however, that
in our model the HOD model is related to the \cmf\ formalism through Eq. (\ref{Phisatb}), 
see Section \ref{sat-gals}. 

As for the observational constraints, in our analysis we include information on the decomposition of the 
\gsmf\ into central and satellite galaxies from robustly constructed galaxy groups; this information is
relevant for constraining the \cmf s \citep[for extensive discussions see][]{Yang+2007,Yang+2008}. 
This is a major difference between our analysis and the one carried out in \citet{Tinker+2013}, where 
weak lensing information from stacking data are used for the observational constraints of their HOD model. 

\section{Observational data}
\label{data}
\begin{figure*}
\vspace*{-70pt}
\hspace*{50pt}
\includegraphics[height=5.5in,width=5.5in]{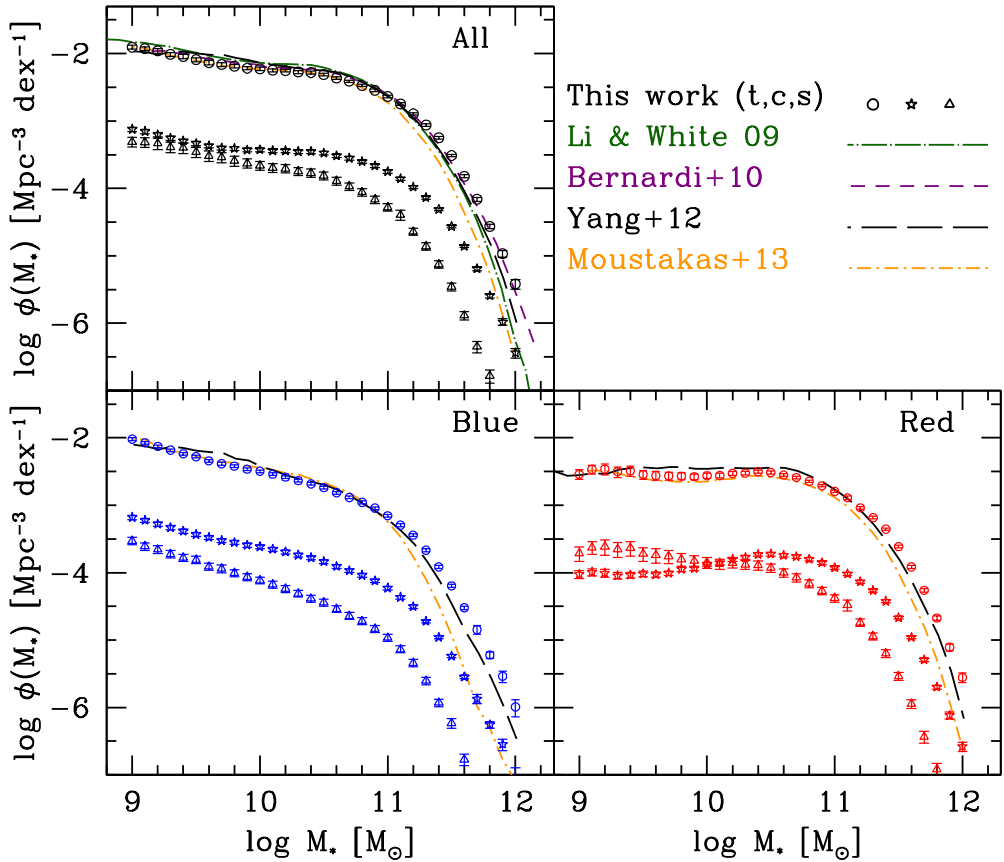}
\caption{  \textit{Upper panel:} The \gsmf\ for all galaxies from the MPA-JHU \nyu/SDSS DR7 sample
obtained and used in this paper (empty circles with error bars) compared with some \gsmf s reported recently in the literature.  
Open stars and triangles with error bars show respectively the
decomposition of the \gsmf\ into
central and satellite galaxies computed by using the \citetalias{Yang+2012} galaxy group catalog. 
The data were shifted down by 1 dex in order to avoid overplotting.  
\textit{Lower panels:} Corresponding blue and red \gsmf s from the MPA-JHU \nyu/SDSS DR7 galaxies,
as well as their decompositions into centrals and satellite galaxies. The orange dot-dashed line is for
the active/passive \gsmf\ decomposition in \citep{Moustakas+2013}.
 }
\label{f1}
\end{figure*}

\begin{table*}
	\caption{Fit parameters to the SDSS DR7 \gsmf s}
	\begin{center}
		\begin{tabular}{l c c c c c c c c c c c c}
			\hline
			\hline
			\gsmf\ &  $\log (\phi_1^*)$ &$\alpha_1$  & 
			$ \log (\mathcal{M}^*_1)$ &  
			$\log (\phi_2^*)$  &$\alpha_2$  &$ \beta $ &  
			$ \log (\mathcal{M}^*_2)$\\
			        (MPA-JHU \ms's) &  [Mpc$^{-3}$ dex$^{-1}$] &  & 
			[\msun] &  
			[Mpc$^{-3}$ dex$^{-1}$]  &  &  &  
			[\msun]\\
			\hline
			All $\ldots\ldots$ & $-2.46\pm0.30$ & $-1.32\pm0.22$  & $9.55\pm0.32$  & $-2.24\pm0.03$  & $-0.64\pm0.15$  & $0.65\pm0.04$  & $10.49\pm0.17$ \\ 
			Blue $\ldots\ldots$ & $-3.07\pm0.13$ & $-1.58\pm0.08$  & $10.17\pm0.13$  & $-2.69\pm0.07$  & $-0.23\pm0.20$  & $0.51\pm0.04$  & $9.82\pm0.19$ \\  		
			Red $\ldots\ldots$ & -- & -- & --  & $-2.60\pm0.02$ & $-0.79\pm0.02$ & $0.80\pm0.02$ & $10.83\pm0.03$\\ 
			\hline
			(Corrected \ms's; see \S\S \ref{mass-correction}) \\ 
			\hline
			All $\ldots\ldots$ & $-3.82\pm0.20$ & $-1.60\pm0.06$  & $11.64\pm0.06$  & $-2.44\pm0.05$  & $-0.15\pm0.12$  & $0.58\pm0.06$  & $10.14\pm0.20$ \\ 
			\hline
		\end{tabular}
		\end{center}
	\label{GSMF-fit}
\end{table*}

As mentioned above, to constrain the model parameters we use  
the observed total blue and red $\gsmf$s 
and the projected \pcf s of blue and red galaxies at various stellar masses. 
In addition, we use the decomposition of the \gsmf s into central and satellites
computed from the \citetalias{Yang+2012} galaxy group catalog.

\subsection{\it Galaxy stellar mass functions} 
\label{gsmfs-obs}

We construct the \gsmf s from the New York Value Added Galaxy Catalogue, \nyu, 
based on the SDSS DR7. We use the sample selection as given in the 
halo-mass based group catalog of \citetalias{Yang+2012}. 
This galaxy group catalog
represents an updated version of \citet{Yang+2007}.\footnote{Available at  
http://gax.shao.ac.cn/data/Group.html.}   
Therefore, our galaxy sample has the same cuts and depurations as in this catalog, 
for further details with respect to the group catalog see  \citet{Yang+2007}. 
The total number of galaxies used for constructing the 
\gsmf\ is 639,359. The \citetalias{Yang+2012} catalog uses colors and magnitudes 
based on the standard SDSS Petrosian radius. 
Following \citet{Blanton+2003b} and \citet{Yang+2009b}, 
we use the evolution correction at $z=0.1$ 
given by $E(z)=1.6(z-0.1)$. For the K-correction, 
we use an analytical model as described in the Appendix. 
In this model, the K-correction term depends on both redshift and 
color, $g-r$, that is, $K=K(z,g-r)$. K-corrections and absolute magnitudes at $z=0.1$ computed 
within this scheme are accurately recovered with typical percentage errors less than 
$\sim10\%$ and $\sim1\%$ on average, respectively.   

For the stellar masses, we use those reported in the MPA-JHU DR7 data base.\footnote{Available at
http://www.mpa-garching.mpg.de/SDSS/DR7.} These masses were calculated from 
photometry-spectral energy distribution fittings assuming a \citet{Chabrier2003} IMF; 
for details, see \citet{Kauffmann+2003}. We found that in our sample
approximately $\approx 9\%$ of the galaxies lack of stellar mass measurements.  
Since this fraction is not negligible, we decided to
calculate the stellar masses for these galaxies by using the color-dependent mass-to-light
ratio given by \citet{Bell+2003}. For this subsample of galaxies we applied a correction of 
$-0.1$ dex in order to be consistent with the \citet{Chabrier2003} IMF adopted in this paper. 
Also, we checked that these galaxies are not particularly
biased in mass or color. The masses of these galaxies were not determined 
likely due to issues in their spectra and/or the stellar populations synthesis fits.

Following \citet{Moustakas+2013}, for the calculation of the
\gsmf\ we adopt a flat stellar mass completeness limit
of $\ms=10^{9}\msun$. As noted by these authors,
this limit is above the surface brightness and 
stellar mass-to-light ratio completeness limits of the SDSS, see
\citet{Blanton+2005,Baldry+2008}.
The \gsmf\ in here is estimated as
\begin{equation}
	\phi_{g}(\ms)=\frac{1}{\Delta\log\ms}\sum_{i=1}^N\frac{\omega_i}{V_{{\rm max},i}},
\end{equation}
where $\omega_i$ is the correction wieight completeness factor in the
\nyu\ and for each galaxy, $V_{\rm max,i}$ is given by
\begin{equation}
	V_{{\rm max},i}=\int_{\Omega}\int_{z_l}^{z_u}\frac{d^2V_c}{dzd\Omega}dzd\Omega.
\end{equation}
Here, $z_l=0.01$ and $z_u={\rm min}(z_{{\rm max},i},0.2)$, $\Omega$ is the solid
angle of the SDSS DR7 and $V_c$ is the comoving volume \citep{Hogg1999}. 
The maximum redshift at which each galaxy can be observed, $z_{{\rm max},i}$,
is computed by solving iteratively the distance modulus equation, i.e.,
\begin{equation}
	m_{{\rm lim},r}-M_{r,i}^{0.1}=5\log D_L(z_{{\rm max},i})-
	25-K(z_{{\rm max},i})+E(z_{{\rm max},i}),
\end{equation}
and each observed galaxy should satisfy the apparent magnitude of the
\citetalias{Yang+2012} group
catalog, $m_{{\rm lim},r}=17.72$. The term $D_L(z)$ is the distance 
modulus \citep{Hogg1999}. 

For each stellar mass bin, errors are computed using the jackknife technique.
We do so by dividing \citetalias{Yang+2012} galaxy group sample into 200 subsamples of approximately equal 
size and each time calculating $\phi_{g,i}(\ms)$. Then, errors are estimated as
\begin{equation}
	\sigma_{\phi}=\left[\frac{N-1}{N}\sum_{i=1}^{N}(\phi_{g,i}-\langle\phi_{g}\rangle)^2\right]^{1/2},
\end{equation}
where $N=200$, $\phi_{g,i}$ is the \gsmf\ of the sample $i$, and $\langle\phi_{g}\rangle$
is the average over the ensemble.  

We divide our galaxy sample into two wide groups, blue and red galaxies. This division
roughly correspond to late-type/star-forming and early-type/passive galaxies, respectively.
Note that for this division, we are using 
K-corrected colors to $z=0.1$, $^{0.1}(g-r)$. Red/blue galaxies are defined
based on the \citet{Li+2006} color-magnitude criteria. These authors separated galaxies 
into blue and red by using a bi-Gaussian fitting model to the color distribution
in many absolute magnitude bins; see \citet{Li+2006} for details.
Because of dust extinction, blue star-forming and highly inclined galaxies  could
be classified as a red passive galaxies \citep[e.g.,][]{Maller+2009}. In \S\S \ref{robustness} 
we discuss the impact of this possible contamination in our color division.

Finally, we also divide our galaxy sample into central and satellite galaxies according 
to the galaxy groups identified in the latest version of the 
\citet{Yang+2007} group catalog. In this paper, we define 
central galaxies as the most massive galaxies within their group.

\subsubsection{\it Total galaxy stellar mass functions} 

Figure \ref{f1} shows the resulting SDSS MPA-JHU DR7 $\gsmf$s for all, blue and 
red galaxies estimated as described above (empty circles with error bars). 
For comparison, in the upper left panel of the same figure we reproduce local
estimations of the total $\gsmf$s based on SDSS 
reported in \citet{Li+2009,Bernardi+2010,Yang+2012} and \citet{Moustakas+2013}. In the 
bottom left and right panels we reproduce the \citetalias{Yang+2012} $\gsmf$s corresponding to
blue and red galaxies, as well as the \citet{Moustakas+2013} $\gsmf$s corresponding to
active and passive galaxies. 
Note that the \gsmf s separated into blue and red
components according to the \citet{Li+2006} color-magnitude criteria are similar
to those of active and passive \gsmf s \citep[from][]{Moustakas+2013} 
for $\ms\lesssim10^{11}\msun$. 

In general, our \nyu/SDSS MPA-JHU DR7 $\gsmf$s for all, blue and red 
galaxies are consistent with previous estimates, except at the
high-mass end, which has a shallower fall than most of previous ones, but in good
agreement with the total \gsmf\ from \citet{Bernardi+2010}.  
In fact, the function could be even shallower if one 
takes into account accurate photometric profile fits
when calculating total luminosities or \ms\ \citep{Bernardi+2013}. In
Section \ref{mass-correction},
we apply corrections to our galaxy stellar masses in order to take into account this 
effect.

As reported by previous authors, we find that a single Schechter function is not consistent
with the total $\gsmf$ \citep[see e.g., ][]{Baldry+2008,Li+2009,Drory+2009,Pozzetti+2010,
Baldry+2012,Moustakas+2013,Bernardi+2013,Tomczak+2014}. 
Besides, the high-mass end of our \gsmf\ is shallower than an exponential decay. 
For completeness, we present the best fits to our \gsmf s. Following \citet{Bernardi+2010}, 
we fit our $\gsmf$s by using a function that is composed of a single Schechter plus a 
Schechter function with a sub-exponential decay at the high-mass end, 
\begin{equation}
	\phi_g(X)dX=\phi_1^*X_1^{\alpha_1}e^{-X_1}dX_1+
		\phi_2^*X_2^{\alpha_2}e^{-X_2^\beta}dX_2,
\label{mod_schec}
\end{equation}
where $X_i=\ms/\mathcal{M}^*_i$ with $i=1,2$. 
The corresponding best-fit parameters 
are reported in Table \ref{GSMF-fit}. 
For blue galaxies we also employed Eq. (\ref{mod_schec}), while for red galaxies we
find that a single Schechter function with a sub-exponential decay gives a good fit to the data. 
The parameters of these fits are also given in Table \ref{GSMF-fit}.

\subsubsection{\it Central and Satellite galaxy stellar mass functions} 

Figure \ref{f1} shows our measurements of the central (stars) and satellite (triangles) 
$\gsmf$s for all, blue and red galaxies. In order to avoid overplotting, we have  
shifted down these $\gsmf$s by $1$ dex. 
The \citetalias{Yang+2012}  galaxy group catalog has been used  
to define central and satellite galaxies; we assume that a central is 
the most massive galaxy in its group.

The blue satellite \gsmf\ lies significantly below the blue central \gsmf\ at all masses. 
In contrast, for red galaxies, at masses below $\ms\sim10^{10}\msun$, the red 
\gsmf\ is dominated by satellite galaxies, roughly by a factor of $\sim2.5$ 
above centrals. For larger masses, the trend inverts and the red \gsmf\ is 
already dominated by red centrals.  When comparing blue and red central $\gsmf$s, 
for $\ms\lesssim10^{10.3}$ \msun, the abundances of red centrals are lower 
than those of blue centrals. For $\ms\gtrsim10^{10.3}$ \msun,  
{\it red centrals become the dominant population}. In the case of satellite galaxies, 
the abundances of blue satellites are lower than those of red satellites
for practically all masses. 

Finally, note that for constraining our model parameters, 
we are not including the full covariance matrix of the different 
$\gsmf$s derived in this work.

\begin{figure}
\vspace*{-35pt}
\hspace*{-10pt}
\includegraphics[height=3.7in,width=3.7in]{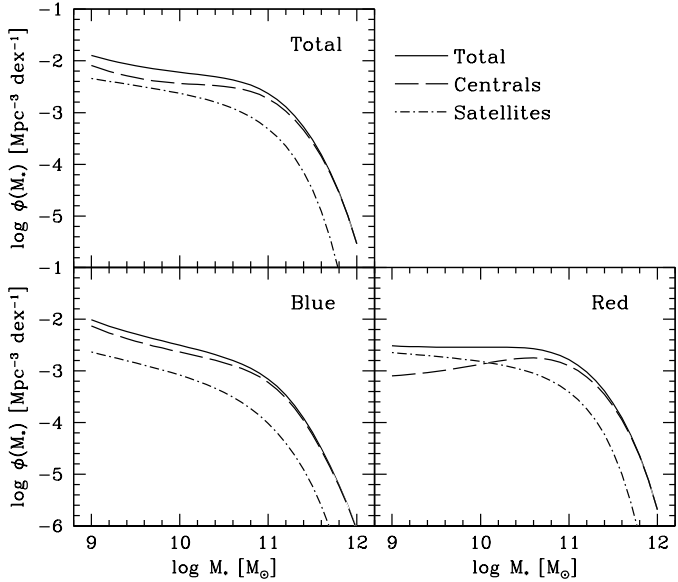}
\caption{Model \gsmf s. Each panel shows the fitted total, blue and red \gsmf s 
(solid lines). The long-dashed and dot-dashed lines show the  
decomposition of the \gsmf s
into central and satellite galaxies, respectively. 
}
\label{f2}
\end{figure}

\subsection{Correlation functions} 

For the correlation functions, we use the \citet{Li+2006} measurements 
of the projected \pcf, $\omega_p(r)$, in five different stellar mass bins 
and for all, blue, and red galaxies. 
This measurements were done based on a sample of $\sim 2\times 10^5$ 
galaxies from the SDSS DR2.
Note that (1) {\it the color-magnitude criterion to separate galaxies into blue and 
red by \citet{Li+2006} is the
same one we have used for constructing the blue and red \gsmf s,} and (2)  
in the calculation of our \gsmf s,  we use the same stellar mass inferences
as in \citet{Li+2006}.

In our analysis, we are using only the diagonal of the covariance matrix
given in \citet{Li+2006}. Unfortunately,  the full covariance matrix is not available
(Cheng Li, private communication).

\begin{figure*}
\vspace*{-180pt}
\hspace*{60pt}
\includegraphics[height=5.3in,width=5.3in]{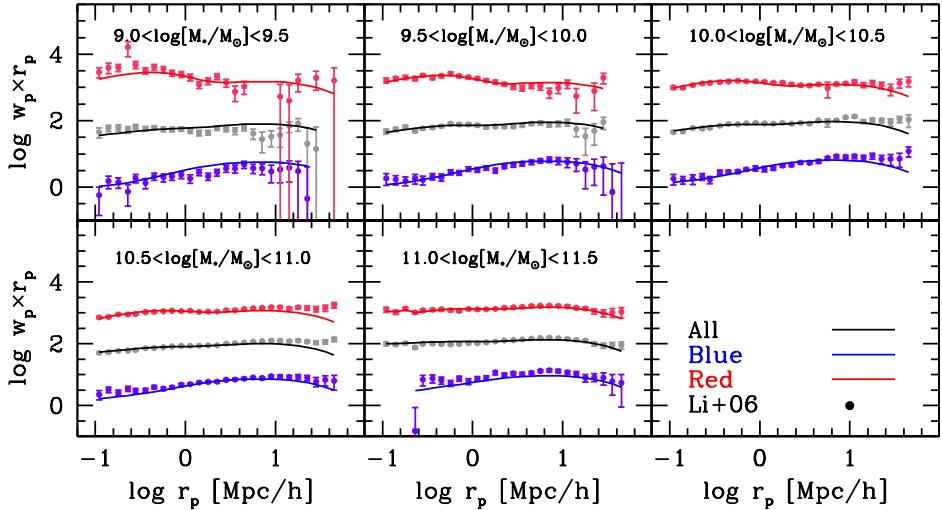}
\caption{Projected \pcf s for all, blue and red galaxies in five different 
stellar mass bins. Best fit models
are shown with solid lines, while measurements based on the SDSS DR2 from \citet{Li+2006} are 
shown with the filled circles with error bars. In order to get a better visual 
comparison between our fits
and observations we plot $\omega_p\times r_p$ instead $\omega_p$. 
}
\label{f2a}
\end{figure*}

\section{Results}
\label{results}

To sample the best fit parameters in our model we run a set of $3\times10^5$  MCMC models.
We obtained the following best fitting parameter:
\\

\begin{itemize}
\item Mean \shmr\ of blue central galaxies (Eqs. \ref{msmh}-\ref{msmh1}):
\begin{equation}
	\begin{array}{c}
		\log\epsilon_b=-1.593  \pm 0.042 \\
		\log M_{1,b}=11.581  \pm 0.034 \\
		\alpha_b=1.500  \pm 0.148 \\
		\delta_b=4.293 \pm 0.271 \\
		\gamma_b=0.396  \pm 0.035 \\
	\end{array}
\end{equation}
\\

\item Mean \shmr\ of red central galaxies (Eqs. \ref{msmh}-\ref{msmh1}):
\begin{equation}
	\begin{array}{c}
		\log\epsilon_r=-2.143  \pm 0.086 \\
		\log M_{1,r}=11.367  \pm 0.100 \\
		\alpha_r=2.858  \pm 0.479\\
		\delta_r=6.026  \pm 0.544 \\
		\gamma_r=0.303  \pm 0.023 \\
	\end{array}
\end{equation}
\\

\item Scatters of the blue/red \shmr s (Eqs. \ref{cmfc},\ref{sigmasj}):
\begin{equation}
	\begin{array}{c}
		\sigma_b=0.118  \pm 0.020 \\
			\sigma_r=0.136  \pm 0.010 \\
	\end{array}
	\label{sigmasbr}
\end{equation}
\\

\item Fraction of blue centrals as a function of \mh\ (Eq. (\ref{fb-Mhalo}):
\begin{equation}
    \begin{array}{c}
          \beta=0.688  \pm 0.065 \\
          b=1.032\pm 0.014\\
     \end{array}
\end{equation}

\item Occupation of satellite galaxies (Eqs. \ref{Phisatb}-\ref{Msatj}):
\begin{equation}
	\begin{array}{c}
		\alpha_{s,b}=-1.251 \pm 0.024 \\
		\phi_{0,b}=-1.324  \pm 0.039 \\
		\phi_{1,b}=0.540  \pm 0.025 \\
		c_0=10.863  \pm 0.037 \\
		c_b=0.192  \pm 0.020 \\
		c_r=0.087  \pm 0.016 \\
		\alpha_{s,r}=-1.096 \pm 0.014 \\
		\phi_{0,r}=-1.363  \pm 0.024 \\
		\phi_{1,r}=1.051  \pm 0.010 \\
	\end{array}
\end{equation}
\end{itemize}

For our best fitting model
we find that the total $\chi^2= 1139$ from a number of $N_d=666$ observational 
data points. Since our model consist of $N_p=23$ free parameters the resulting 
reduced 
$\chi^2$ is $\chi^2/{\rm d.o.f.}= 
1.77$.   Figure \ref{PD1} in Appendix \ref{PD} shows the posterior probability distributions 
of the model parameters.  This plot gives a visual information of the covariances between the
model parameters. 
In almost all the cases, the parameters do not correlate between each other.

\subsection{\gsmf s and correlation functions}
\label{gsmf_wp}

In each panel of Fig. \ref{f2} we plot the best-fit model \gsmf s 
for all, blue and red galaxies (solid lines). 
In the same panels, we show the decomposition of
the $\gsmf$s into central (long-dashed lines) and satellites (dot-short-dashed lines) 
corresponding to all, blue and red galaxies. In general, our model fits describe well  
the observed \gsmf s, decomposed into central and satellites (compare Fig. \ref{f2} with  
Fig. \ref{f1}).

Figure \ref{f2a} shows the observed projected $\pcf$s reported in 
\citet[][filled circles with error bars]{Li+2006} and the best model
fits (solid lines). The projected $\pcf$s are for all, 
blue, and red galaxies (black, blue and red colors, respectively)
in five different stellar mass bins. In order to get a better visual 
comparisons of our fits to observations, 
we plot $\omega_p\times r_p$ (instead of $\omega_p$) as a function of $r_p$. 
For clarity, our fits and the data points for blue and red galaxies 
have been shifted by +1 dex and -1 dex,
respectively. In general, our fits describe well the observations 
for all mass bins and in almost all separations. We note, however, that at
large separations our fits tend to lie below observations in the stellar
mass bins $10<\log(\ms/\msun)<10.5$ and $10.5<\log(\ms/\msun)<11$. It is known that
at large separations the correlation functions are affected by cosmic variance in
volume-limited samples. This effect in the SDSS galaxies has been investigated in 
detail by \citet{Zehavi+2005} and \citet{Zehavi+2011}, where the authors find
that the most significant cosmic variance effect appears due to the 
presence of a supercluster at $z\sim0.08$, the Sloan Great Wall \citep{Gott+2005}. 
These authors conclude that the inclusion of
this supercluster in the calculation of the correlation functions 
causes an anomalous high amplitude of galaxy clustering at large 
separations in samples with $-21<M_{r,0.1}<-20$. This magnitude bin
roughly corresponds namely to a stellar mass bin $10<\log(\ms/\msun)<11$.
Thus, the observational values of $\omega_p$ at large radii in this mass bin
could be overestimated.

\begin{figure}
\vspace*{-190pt}
\hspace*{0pt}
\includegraphics[height=7.5in,width=6.8in]{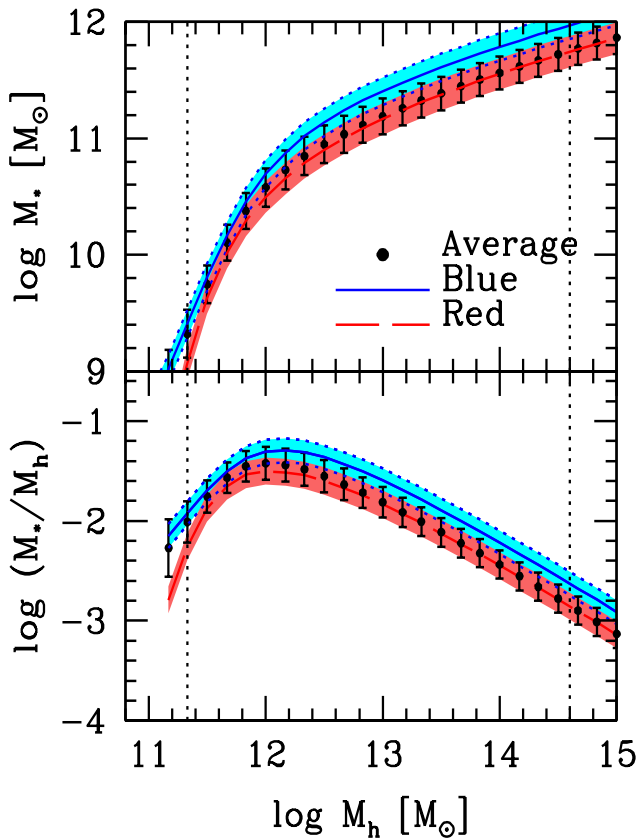}
\caption{ \textit{Upper panel:} Inferred \shmr s of blue and red central galaxies 
(blue and solid lines, respectively) and 
the density-averaged \shmr\ (dots). The shaded areas and error bars are their
corresponding constrained scatters ($\sigma_j$), which can be considered   
as upper limits to the intrinsic scatters.
\textit{Bottom panel:} The \ms-to-\mh\ ratio as a function of
\mh\ for the same cases plotted in the top panel. 
Note that the density-averaged \shmr\ approaches the red \shmr\ at large masses,
while at lower masses is in between the blue and red \shmr s.  
The dotted vertical lines indicate the lower and upper limits in halo mass at which our 
average $\shmr$ can be robustly constrained. 
}
\label{f3}
\end{figure}

\begin{figure*}
\vspace*{-270pt}
\hspace*{50pt}
\includegraphics[height=5.8in,width=5.8in]{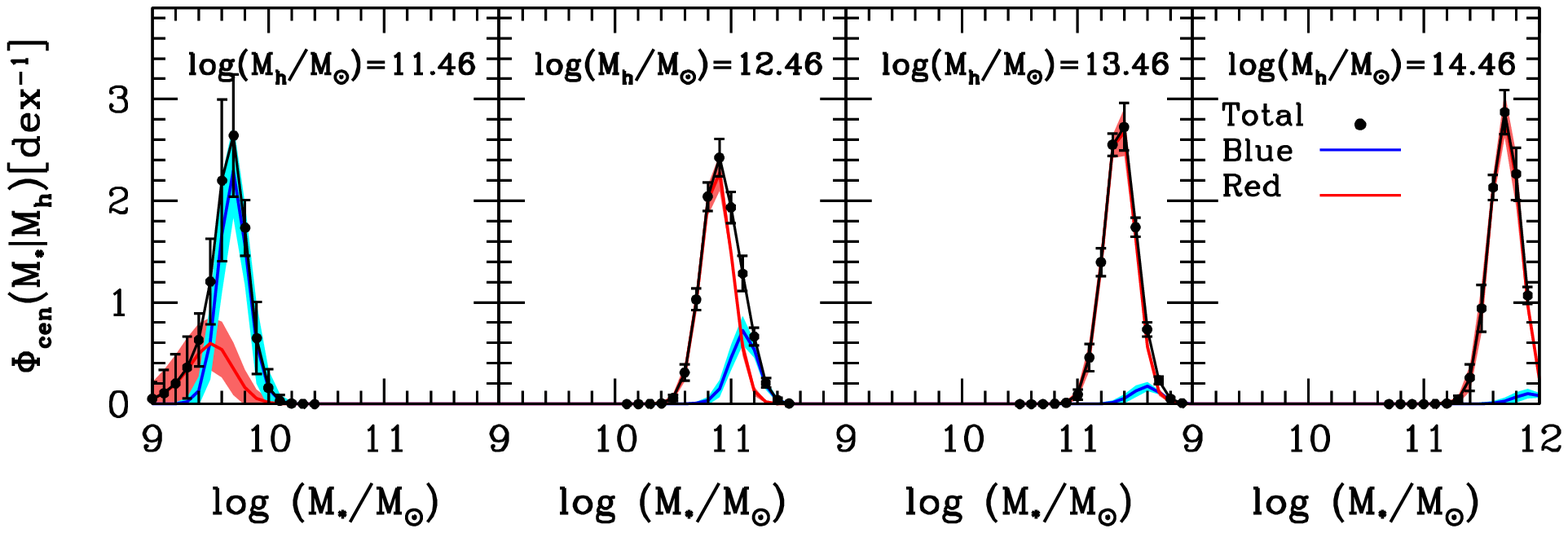}
\caption{Inferred blue and red central conditional stellar mass distributions at different 
halo masses (red and blue lines).
The shaded areas are the $1\sigma$ error due to the uncertainties in the model parameters. 
The dotted points with error bars show the total central conditional distributions. Note the there is
a bimodality at low halo masses, while at large masses the total distribution is dominated by red
central galaxies. 
}
\label{f11}
\end{figure*}

In order to quantify the effect of cosmic variance in our fits,
we have excluded separations larger than $r_p>10 h^{-1}$ Mpc in the projected \pcf s in the 
bins that encompass masses in the range $10<\log(\ms/\msun)<11$  and recalculated 
the $\chi^2/$d.o.f by using our best fit parameters. The exclusion of these separations leads to  
a substantial decrease in the reduced $\chi^2$, $\chi^2/$d.o.f.$=1.47$. 
This implies that cosmic variance effects could affect our fits. 
To test this, we have recalculated our model parameters using this modified data set regarding the \pcf s. 
As a result, we found that the values of all the model parameters remain similar to those 
obtained previously, particularly those related to the \shmr s. 
The insensitivity of HOD model parameters to the data at large separations has been
reported previously \citep[see e.g.,][]{Zehavi+2011}. 
The reason is that for a fixed cosmology, the
HOD parameters have less freedom to adjust the large-scale
correlation relative to the (more) robust inferences at small separations, $r_p<2h^{-1}$Mpc. 
Therefore, we conclude that our model fit parameters are robust against uncertainties
due to cosmic variance in the \pcf s at large separations.  

On the other hand, in the stellar mass bin $9<\log(\ms/\msun)<9.5$ we note that
our fits tend to be slightly below observations at small separations. 
It is not clear the reasons behind this difference but
we have quantified the impact of it in our goodness-of-fit. Similarly done for large
separations, we have recalculated the $\chi^2/$d.o.f. but this time excluding the information
of the correlation functions at this mass bin. The resulting 
$\chi^2/$d.o.f. is 1.71. This means that the correlation function from this stellar mass
bin does not add substantial information to constrain our model parameters. Note, however,
that in order to obtain robust inferences of our model parameters, we have also included the
information of the central and satellite $\gsmf$s.

\subsection{Stellar-to-Halo Mass Relations}
\label{SHMRs}

The derivation of the \shmr s and their scatters for local blue and
red central galaxies is the main goal of this paper. These mass
relations and the corresponding \ms-to-\mh\ ratios vs. \mh, as constrained by means of 
our model, are shown in the upper and lower panels of Fig. \ref{f3}. 
Blue solid and red long-dashed lines show the mean relations of blue and red centrals,
respectively. Shaded areas show the standard deviations of the (lognormal-distributed) 
scatter (see Section \ref{central-galaxies} and Eq. \ref{cmfc}). Recall that this scatter
is composed by an intrinsic component and by a measurement error component 
(Eq. \ref{sigmasj}). This scatter should be considered as an upper limit of the intrinsic 
component. 
In Section \ref{scatter}, we estimate values for the intrinsic scatters 
once we introduce conservative estimations for measurement error components.
Black dots with error bars represent the average central relations, 
calculated as in Eq. (\ref{mean_shmr}),
and its corresponding scatter (standard deviation, 
calculated using Eq. \ref{sgima_msmh}).
Recall that Eq. (\ref{mean_shmr})
is a {\it density-weighted average}, so the mean $\shmr$ for all central 
galaxies is located in between the $\shmr$s of blue and red centrals. 
However, at high masses the average $\shmr$ is practically equal to the one of
red galaxies. This is because almost all massive halos host red central galaxies, see below.
The dotted vertical line indicates the lower limit in halo mass at which our 
average $\shmr$ has been robustly constrained. 
Note that we did not assume any functional form both for the shape and for the scatter of 
the average \shmr. Instead, they are a direct {\it prediction} and the 
result of constraining separately the $\shmr$s of blue and red central galaxies. 

Our results point out that the \shmr s of blue and red centrals are different. In other words,
{\it the \shmr\ of central galaxies is segregated by color}. For a given 
\mh, the \ms\ or \ms-to-\mh\ ratio of blue galaxies is always larger than the one 
of red centrals. 
The minimum difference is of 0.16 dex at $M_{\rm h,min}=5\times 10^{11}$ \msun,     
close to the mass where the halos contain the same fraction of blue and red centrals 
(see below, \S\S \ref{blue-halos}). The difference increases up to $\approx 0.24$ dex at 
$\mh=5\times 10^{12}$ \msun\ and for larger masses it remains roughly constant. 
For masses $<M_{\rm h,min}$, the difference strongly increases.  
At any mass, the differences are larger than the $1\sigma$ scatter of the relations, 
$\sigma_b$ and $\sigma_r$, respectively (see also \S\S \ref{scatter} below).  
This can be also appreciated in Fig. \ref{f11}, where we plot the conditional stellar 
mass distributions of blue and red central galaxies for four different halo masses, 
blue and red lines, respectively. The shaded areas show the 
{\bf $1\sigma$ confidence intervals} 
around these relations.  The dots with error bars show the same but for all central galaxies. 
The distributions of blue and red centrals, which are related to the scatters 
around the corresponding \shmr s, are different.
Besides, as it will be shown in the next subsection and in Fig. \ref{fig-scatter}, the 
scatters around the constrained blue and red \shmr s 
compared to the $1\sigma$ confidence intervals obtained from our MCMC run
are smaller than the corresponding scatters. Thus, the differences in the values
of the mean \shmr s of blue and red centrals are significant also at the level of error analysis. 

Figure \ref{f11} shows that for $\mh\lesssim 3\times 10^{11}$ \msun, 
the galaxy population of centrals  is dominated by blue galaxies over the 
red ones, while at $\mh\sim 3\times 10^{12}$ \msun, red galaxies are already 
the dominant population but there is yet a non-negligible fraction of blue galaxies.
Instead, at $\mh\sim 3\times 10^{13}$ \msun, practically all central galaxies are red. 
{\it It is at masses $10^{11.5}\msun\lesssim\mh\lesssim10^{12.5}\msun$ that
the \shmr\ of central galaxies has a clear bimodal distribution.}  
At larger and lower masses, the average \shmr\ for all central galaxies is 
practically given by the \shmr\ of red and blue centrals, respectively (Fig. \ref{f3}).    

 \subsubsection{The  scatter of the $\shmr$s}
 \label{scatter}
 
In our model, we have assumed that the stellar mass distribution of both
 blue and red central galaxies are lognormal. Additionally, we assumed that the widths (scatters)
of these distributions are free parameters in our model. Recall, that these scatters 
 were assumed independent of \mh\ (Eq. \ref{cmfc}). 
The values constrained for these scatters are $\sigma_b=0.118\pm0.020$ dex and 
$\sigma_r=0.136\pm0.010$ dex, respectively (solid blue and short-dashed red lines 
surrounded by shaded areas in Fig. \ref{f3}). In other words, the scatter around the 
\shmr s of red and blue central galaxies is higher for the former than for the latter, 
though the error bars overlap slightly. This implies that the \shmr\ of blue central 
galaxies is tighter than the one of red centrals.

\begin{figure}
\vspace*{-270pt}
\hspace*{0pt}
\includegraphics[height=6.5in,width=6.5in]{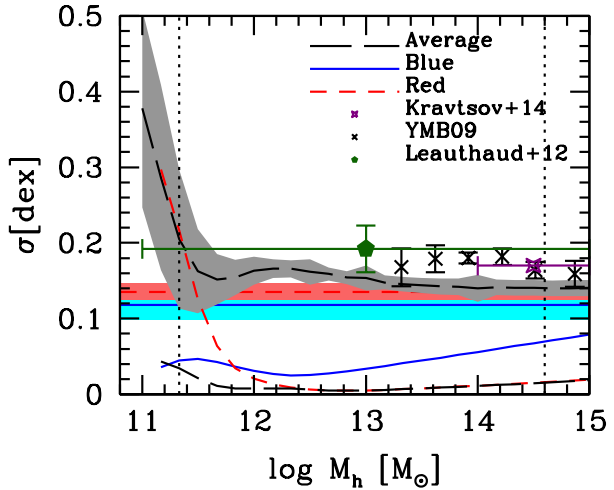}
\caption{Different type of scatters around the \shmr s.
Blue solid, red short-dashed and black long-dashed lines are the 
constrained scatters, $\sigma_b$, $\sigma_r$ and
$\sigma_A$, which can be considered as upper limits to the
respective {\it intrinsic} scatters (see text); the shaded areas are 
the corresponding $1\sigma$ confidence intervals.  Note that $\sigma_A$ 
changes with mass, while $\sigma_b$ and $\sigma_r$ were assumed to be constant. 
The bottom blue solid, red short-dashed and black long-dashed lines 
are the magnitude of the $1\sigma$ confidence intervals from 
$3\times 10^5$ MCMC trials for the blue, 
red and average \shmr s, respectively. Additionally, we have plotted 
the scatter constrained by \citet[][skeletal symbols]{Yang+2009b}, 
\citet[][filled pentagon]{Leauthaud+2012},
and \citet[][open star]{Kravtsov+2014}. 
 The dotted vertical lines indicates the lower and upper limits in halo mass at 
which the intrinsic scatter $\sigma_A$ can be robustly constrained. 
}
\label{fig-scatter}
\end{figure}

The scatter around the density-averaged (blue + red) \shmr, $\sigma_A$, is plotted
in Fig. \ref{fig-scatter} (black long-dashed line) along with $1\sigma$ confidence interval (gray shaded area). 
Similarly to Fig. \ref{f3}, the dotted vertical line indicates the lower limit in halo mass at which the 
scatter $\sigma_A$ has been robustly constrained. 
For $\mh\sim10^{11.3}$ to $\sim10^{15}$ \msun, $\sigma_A$ changes from
$\sim0.20$ dex to $\sim0.14$ dex. 
Note that $\sigma_A$ is constant for masses above
$\mh\sim10^{13}$ \msun. The dependence of $\sigma_A$ on \mh\
naturally arises according to Eq. (\ref{pcen}). Because the majority of high mass halos,
$\mh\grtsim10^{13.5}$ \msun, host red central galaxies (see Fig. \ref{f11}; see also Fig. \ref{f9} below)
then $\sigma_A\sim\sigma_r\approx 0.14$ dex. In contrast,
the increasing of $\sigma_A$ with decreasing \mh\ is the result of the color bimodality 
in the average \shmr\ for masses below $\mh\lesssim10^{12.5}$ \msun\ (see Fig. \ref{f11}).

As mentioned in \S\S \ref{central-galaxies}, the scatters $\sigma_{j}$ reported
above consist of an intrinsic component, $\sigma_{j}^{\rm i}$, and of  
measurement errors component, $\sigma_{j}^{\rm e}$, see Eq. (\ref{sigmasj}).
Following the results by \citet{Behroozi+2010} and \citet{Leauthaud+2012}, we assume that the    
dominant source of measurement error comes from individual stellar mass estimates. 
For the data employed in our analysis, we used stellar masses from an update 
of \citet{Kauffmann+2003}. In that paper, the authors found that the 95 per cent confidence 
interval from their stellar mass estimates is typically a factor of $\sim2$ in mass. As noted
by the authors, in a Gaussian distribution this corresponds to four times the 
standard error, resulting in a standard deviation of $\sim0.075$ dex. 
This is consistent with the standard deviation reported in \citet{Conroy+2009b} for
local luminous red galaxies. 
Measurement errors might be different between red and blue galaxies, for example 
\citet{Kauffmann+2003} found that these errors are somewhat smaller
for older galaxies with ${\rm D}_n(4000)>1.8$ than for younger galaxies with  
${\rm D}_n(4000)<1.8$. Unfortunately, the authors do not provide values for their 
confidence intervals. 

We assume a conservative value of $\sigma_{j}^{\rm e}=0.07$ dex for both 
blue and red central galaxies. We estimate the intrinsic scatters of blue
and red central galaxies by deconvolving from measurement errors using the 
$3\times 10^5$ MCMC models generated for fitting the data. We estimate conservatives
values of $\sigma_{b}^{\rm i}=0.094\pm0.024$ and $\sigma_{r}^{\rm i}=0.116\pm0.012$ 
for blue and red centrals, respectively.

Finally, in Fig. \ref{fig-scatter}, we also plot the magnitude of the $1\sigma$ confidence intervals 
around the mean $\shmr$s for blue, red and all central galaxies (solid blue, short-dashed red, 
and long-dashed black lines without shaded areas, 
respectively). These confidence intervals were obtained from the $3\times10^5$ 
MCMC models generated to sample the best fit parameters in our model. 
The confidence intervals take into account the error bars from 
our set of observational constraints. 
Note that we are not taking into account any source of systematical uncertainty, which
are usually dominated by systematical uncertainties in stellar mass inferences and they may 
be up to $0.25$ dex \citep[][]{Behroozi+2010}. 
The confidence intervals around the mean 
$\shmr$ of all central galaxies (black long-dashed line) 
are much smaller than the scatter $\sigma_A$ (black long-dashed line      
surrounded by a gray area). Similarly, the confidence intervals
around the mean $\shmr$s of blue and 
red central galaxies are much lower than their scatters \sigmaj, 
excepting at low and high masses for red and blue galaxies, respectively. 
At these masses, the $1\sigma$ confidence intervals around the mean $\shmr$s
increase due to the scarce data and large observational error bars in these limits. 

\begin{figure}
\vspace*{-110pt}
\includegraphics[height=6.5in,width=6.5in]{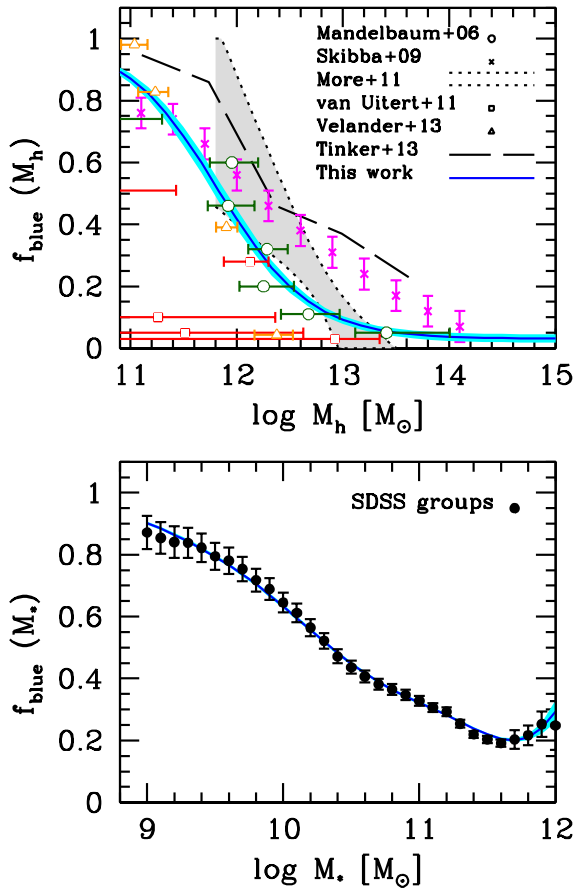}
\caption{ \textit{Upper panel:}  Inferred fraction of halos hosting central blue and red galaxies 
as a function of \mh, blue and
red solid lines, respectively. The blue and red shaded areas correspond to the $1\sigma$ 
confidence intervals. 
Most of the symbols with error bars are observational inferences using galaxy-galaxy weak 
lensing, see text. Long dashed line show the best fitting model from \citet{Tinker+2013} 
based on the combined analysis of galaxy weak-lensing and clustering from COSMOS 
field at $z=0.36$. 
Skeletal symbols show the fraction of blue centrals from \citet{Skibba+2009} model.  
\textit{Bottom panel:} The same fraction of central blue galaxies as in the upper panel 
but as a function of \ms. Solid circles
with error bars are results from the \citetalias{Yang+2012} galaxy group catalog.
}
\label{f9}
\end{figure}

\subsection{The fraction of halos hosting blue/red central galaxies}
\label{blue-halos}

\begin{figure*}
\vspace*{-180pt}
\hspace*{50pt}
\includegraphics[height=5.5in,width=5.5in]{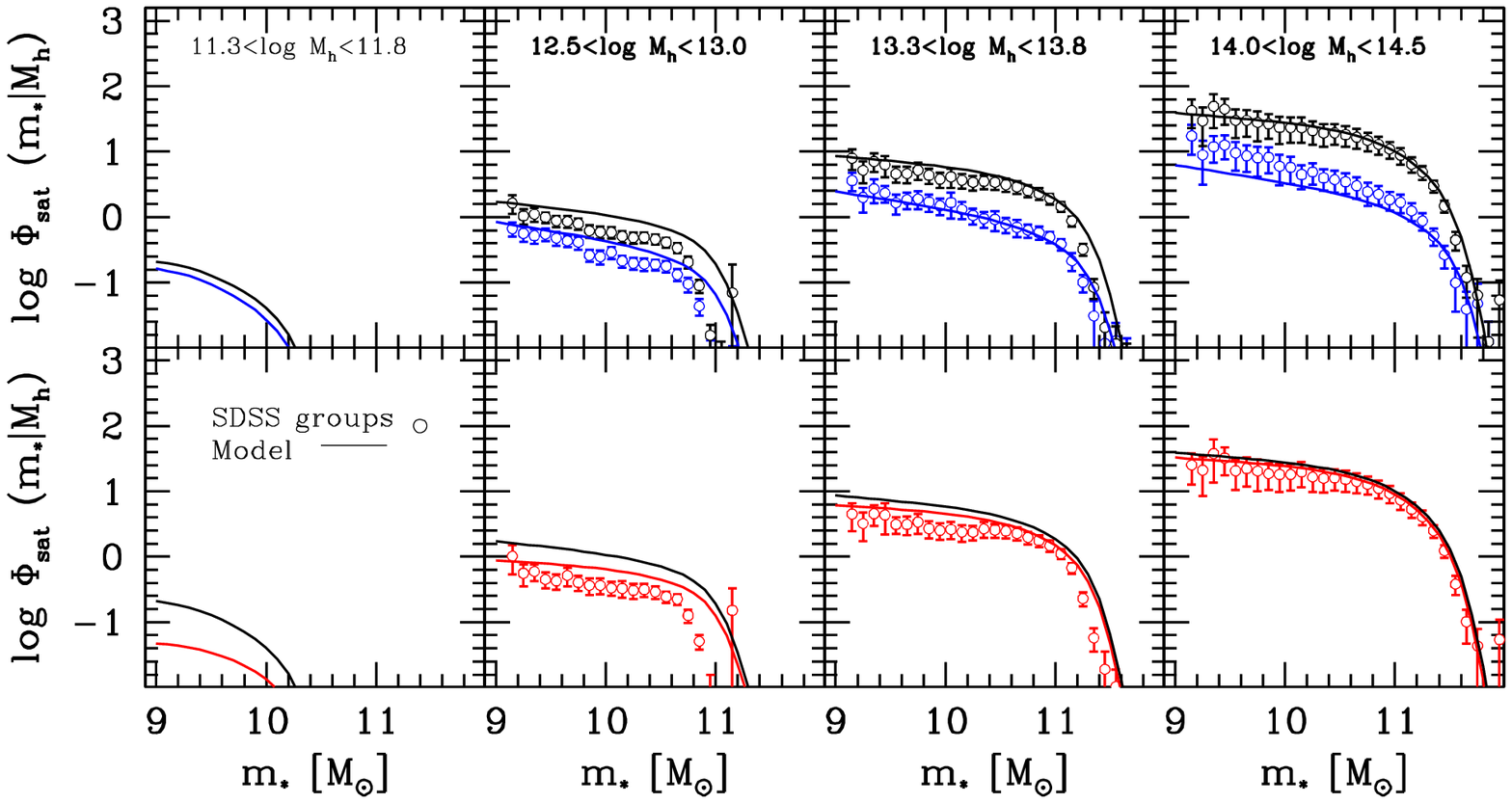}
\caption{{\it Upper Panels:} Satellite \cmf s of all and blue galaxies 
for four different halo mass bins (indicated inside the panels). Solid lines
are for the model predictions and open circles with error bars are from
 \citetalias{Yang+2012} group catalog; black and blue colors are for all
 and only blue satellites, respectively.
{\it Lower Panels:} As in in the upper panel but for red satellite galaxies only. 
For comparison black solid lines reproduce our best fitting model for the satellite \cmf s of all galaxies.
}
\label{f10}
\end{figure*}

We have parametrized the fraction of halos hosting blue/red (approximately active/quenched) 
central galaxies by using a general observationally-motivated dependence, see \S\S \ref{central-galaxies}.
In Fig. \ref{f9}, we show  the constrained blue fraction as a function    
of \mh, $\fb(\mh)$. Recall that \fb\ is the complement of \fr, $\fb(\mh)= 1 - \fr(\mh)$. 
The fraction of blue centrals decreases with \mh\
roughly a factor of $\sim4.4$ from $\mh \approx 10^{11}$ to $\approx 10^{12.5}$ \msun. 

Our results show that $b\approx1$ in Eq. (\ref{fb-Mhalo}). This means that the dependence 
of \fr\ on \mh\ is actually close to the one suggested in \citet{Woo+2013} based on the 
empirical inferences of \citet{Peng+2012}. In this case, the characteristic halo mass 
where the fraction of halos hosting blue and red centrals 
is the same, i.e., $f_{\rm blue}=f_{\rm red}=0.5$ is given by: 
$\mh= 6.88\pm0.65\times 10^{11}$\msun.
For Milky-Way sized halos, $\mh\sim10^{12}\msun$, 
the blue (red) fraction is $\sim 1/3$ ($\sim 2/3$). 

In Fig. \ref{f9}, we reproduce also some direct observational inferences obtained
by stacking weak-lensing data: \citet[][open circles]{Mandelbaum+2006},
\citet[][open squares]{vanUitert+2011} and \citet[][open triangles]{Velander+2014}.
In general, our result is in reasonable agreement with these direct inferences. 
We also compared our resulting fraction with the analysis of satellite kinematics performed in 
\citet[][gray shaded area]{More+2011}. The shift observed with respect to our results is possible   
related to the fact that halo masses measured from satellite kinematics are usually 
higher than other methods, see Section \ref{comparison}.  
We also compare our results with those of \citet[long dashed-line][]{Tinker+2013}.
Note, however, that the blue fraction from \citet{Tinker+2013} is higher that for local
galaxies. One possibility is that the fraction of halos hosting blue centrals is expected 
to be higher than for $z\sim 0$. Another reason is that
due to sample variance in the COSMOS field at $z\sim0.36$, it may not possible to obtain
a direct comparison with our results (Jeremy Tinker, private communication). 
We also include a comparison with results from \citet{Skibba+2009}. Although at low masses 
their results are consistent with our blue fraction, at the high mass end there is a discrepancy. 
This is possible related to the fact the the  \citet{Skibba+2009} model assumes that color depends 
only weakly on mass. 
Finally, it is important to note, that the way the two galaxy populations are separated in all the 
studies mentioned above, may vary with most of the authors (based on colors,
specific star formation rate, type, etc.). 
This introduces differences in the fractions plotted in Fig. \ref{f9}. In spite of that,
the overall result is consistent among most of the studies.

For completeness, the lower panel of Fig. \ref{f9} shows our model fitting 
(solid line with a shaded area) to the empirical fraction of blue central galaxies as a function of 
\ms, i.e. the ratio of blue-to-all central \gsmf s from the \citetalias{Yang+2012} galaxy
group catalog (solid circles with error bars). 

\subsection{The blue/red satellite \cmf s}
\label{satellite-fractions}

Figure \ref{f10} shows the predicted \cmf s for all, blue and red satellites 
(Eqs. \ref{Phisatb}--\ref{Msatj}). For comparison, the circles with error bars in all
the panels show the same but for the \citetalias{Yang+2012} galaxy group catalog 
(we use only those groups that are complete according to 
the completeness limits discussed in Section 2.2 of \citet{Yang+2009b} for both
halo and stellar masses). We have corrected their halo masses to match our 
virial halo mass definition, see Eq. (\ref{m200}) in Appendix \ref{halo-mass}. 
Error bars were estimated by using the jackknife method described in 
Section \ref{data} with $N=200$.
We have also weighted each galaxy by a factor of $V/V_{\rm max}$. 
In general, both the amplitude and the shape of the predicted \cmf s agree 
with those from the Y12 group catalog. In more detail, for massive halos, 
the predicted \cmf s for blue satellites are shallower in 
the low-mass end ($\ms\approx10^{10}\msun$) than the inferred ones from the group catalogs
but within the error bars. One possible reason is the relative poor constraints from
the lowest mass bin from $\omega_p$, see the discussion in Section \ref{gsmf_wp}.


\section{Discussion}
\label{Discussion}

\subsection{Robustness of the results}
\label{robustness}

The segregation in color found in the \shmr\ of central galaxies in 
Section \ref{results} is a relevant result. How robust is it?  We have carried
out several experiments in order to explore this question.  For example,
we have explored what happens if we force the \shmr\ of blue and red
centrals as well as their scatter to be identical. 
In this case, we find that the best fits to observations are poorer than those
obtained in Section \ref{results}, with $\chi^2/$d.o.f.$=2.1$, which is $\sim 25\%$ 
larger than in that Section. 
We have also checked what fraction of our $3\times10^5$ MCMC models
have similar parameters in their parametrization for the blue and red \shmr s. 
First, note that most of the error bars in each model parameter is of the order 
of $\sim10\%$ of the value of the parameter. When allowing this relative difference 
of $\sim10\%$ between models we could not found any model. By
allowing for relative differences
in the parameters up to a 50\%, only a $\approx 3\%$ of the MCMC models obey
this condition. Therefore, in the search of the best fits, the cases of similar
blue and red \shmr s are really rare.  

In addition, we have found (as expected) that the key ingredient of the color 
segregation in the \shmr\ is the fraction of halos hosting red (blue) centrals as a function of \mh. 
This fraction, according to our results, is essentially defined by the parameter, 
\mhchar\ (see Eq. \ref{fb-Mhalo}) and in less degree by the parameter $b$. 
Under the assumption that {\it the \shmr\ of blue and red centrals are identical} 
and keeping all the parameters for satellite galaxies as obtained in Section \ref{results}, 
excepting \mhchar, we find that the best fits that minimize $\chi^2$ is with 
$\mhchar\approx 9.7\times10^{11}$ \msun, i.e., $\sim1.4$ times larger than the one found 
in Section \ref{results}. The fits to observations in this cases are very poor, 
giving a total $\chi^2/$d.o.f$\approx 4.7$. This difference shows that the segregation 
in color is sensitive to the value of \mhchar. If now \mhchar\ is fixed to larger 
values than $\sim 10^{12}$ \msun, and one allows for a difference between the \shmr\ of 
blue and red galaxies, then we obtain that the constrained \shmr s differ in an opposite way 
as found in Section \ref{results} (obviously, the fits become even much poorer). On the other 
hand, as \mhchar\ becomes smaller, the difference between the \shmr s in the direction 
of blue galaxies gets larger \ms\ for a given \mh\ than red ones. 

We also explored the case of generalizing the function Eq. (\ref{fb-Mhalo}) by 
allowing the mass term in the denominator to vary as $(\mhchar/\mh)^a$. We have found that 
the best fit is obtained when $a\approx 1$, that is, the proposed function 
Eq. (\ref{fb-Mhalo}) is robust. 
The resulting \shmr s in this case are very close to those obtained in Section 
\ref{results}, where $a=1$ is assumed.
Along the same vein, if we modify partially our
parametric approach, then the obtained \shmr s do not change significantly, though
other predictions may already differ. For instance, this
was the case when we modeled the satellite \cmf s through the \amt\ between the theoretical 
subhalo mass function and the (predicted) satellite \gsmf\ (case B in RAD13),
instead of proposing parametric functions for the  \cmf s.

In our model, we assumed $\sigma_b$ and $\sigma_r$ as free parameters. 
The resulting constrained values were $\sim 0.11$ and $\sim 0.14$, respectively.
On the other hand, these parameters can be also obtained directly from the group 
galaxy catalogs, especially for massive halos \citet[see e.g.,][]{Yang+2009b}, which
are slightly different form our model constraints, see Section \ref{scatter-discussion} below. 
To test the impact of this,
we repeat our analysis but this time we fix $\sigma_b=\sigma_r=0.15$ dex. As we 
describe in Section \ref{scatter-discussion}, $\sigma_b\approx \sigma_r\approx 0.15$ dex   
as derived directly from the \citet{Yang+2007} galaxy group catalog.                                      
In this case, we obtain a reduced $\chi^2$ similar to the one obtained
in Section \ref{results} (eq. \ref{sigmasbr}), where we left  $\sigma_b$ and $\sigma_r$ 
as free parameters. 
It should be said that the blue and red \shmr s obtained in the former case results slightly 
shallower at the  high-mass end than those reported in Section \ref{results}. However, the relative 
separation between these two relations remains roughly the same, showing that our 
result that the \shmr\ segregates significantly by color is robust to changes
in the $\sigma_b$ and $\sigma_r$ parameters.

Finally, we have explored the sensitivity of the blue/red central \shmr s to the observational
constraints. In particular, in one experiment we renounced to use as constraints 
the central/satellite \gsmf\ decompositions based on the \citetalias{Yang+2012} group catalog,
only the total blue/red \gsmf s (and the \pcf s) were used. 
Again, the constrained blue, red and average central \shmr s, remained almost the same. In this
experiment, the central/satellite \gsmf s are predicted; they are in reasonable agreement with those
obtained with the \citetalias{Yang+2012} group catalog, excepting the low-mass end of the blue satellite
\gsmf, which also implied a poor agreement with the \citetalias{Yang+2012} \cmf s of blue satellites
at low stellar masses in halos smaller than $\sim 10^{13}$ \msun.

 \subsubsection{On the blue/red separation criterion}
 
In this paper, central galaxies were separated into blue and red galaxies
by using the color-magnitude criterion of \citet{Li+2006}. While this separation is very rough,
some of the results discussed in Section \ref{results} could be sensitive to it. 
It is well known that there is not a perfect correspondence between
blue/red galaxies and disk-/bulge-dominated or active/passive ones 
\citep[c.f.][]{Maller+2009,Bundy+2010,Woo+2013}.
Late-type (blue) galaxies can appear in our separation as early-type (red) galaxies
if they color is red due to dust extinction, specially when they are highly inclined and massive. 

In order to evaluate the sensitivity of our results to dust extinction effects, we compare 
our fraction of blue (red) galaxies as a function of absolute magnitude in the $r$-band 
at $z=0.1$, $^{\rm 0.1}M_{r}$, with the one presented in \citet{Jin+2014}.
These authors analyzed a subsample of the SDSS DR7 galaxies by selecting 
face-on galaxies only. By means of the $(u-r)_{\rm 0.1}$  color-magnitude diagram, they 
separate their sample into blue, red, and green galaxies, the latter are actually a small 
fraction ($<15\%$).  The $^{\rm 0.1}M_{r}$ at which the fraction of red and blue galaxies 
is equal is $\approx -20$ mag both in \citet{Jin+2014} and in our case. 
At $^{\rm 0.1}M_{r}=-21.4$ mag, which is the highest magnitude in the 
\citet{Jin+2014} sample, their ratio of red to blue 
galaxies is $\sim 1.40$ while in our case this ratio is $\sim 1.56$, that is, their face-on 
sample of galaxies contains only $\sim 11$\% less red luminous galaxies than our one.  
At $^{\rm 0.1}M_{r}=-21$ mag, 
the situation inverts, i.e., their sample contains more red galaxies than ours. 
We conclude that dust extinction does not affect significantly our rough separation between
blue and red galaxies and its corresponding identification with late and early types, respectively.

\subsection{Interpretation of the results}

The \ms-to-\mh\ ratio is commonly interpreted as 
the efficiency of galaxy stellar mass growth (mainly by in situ star formation) within 
dark matter halos. Our semi-empirical results show that {\it the \ms-to-\mh\ 
ratio of central galaxies is segregated by color at all masses}, with blue galaxies having
higher ratios than red ones (Fig. \ref{f3}). This could be interpreted as that blue central 
galaxies were more efficient in assembling their stellar masses than red centrals 
at a given halo mass. However, the \ms-to-\mh\ ratio is actually a time-integrated 
quantity, result of the combination of two aspects: 
\begin{enumerate}
\item The efficiency of stellar mass growth, both by star formation (SF)  
 in situ and by the accretion of satellites (mergers).

\item The processes that halt galaxy growth (particularly due to quenching of the SF) 
at a given epoch, while the halo mass continues growing. 
\end{enumerate}
In order to explain the fact that  (1) blue centrals have higher \ms-to-\mh\ ratios than 
red centrals (Fig. \ref{f3}), and (2) that low (high) mass halos 
are dominated by blue (red) centrals (Figs. \ref{f11} and \ref{f9}), we need to understand 
which of the above mentioned process have played a dominant role.
Our results suggest that it should be the second one, i.e., the {\it galaxy 
quenching} process.\footnote{Galaxy quenching is commonly thought
as a process of SF rate fading rapidly from actively star forming to quiescent. In this sense, 
it is possible that a recently quenched galaxy (low specific SF rate) is yet blue.  
We use the concept of quenching in a general way, assuming that once a galaxy becomes red
it is quenched and it will remains so.}

In the context of the quenching scenario, while the SF in a galaxy is ceased, its host \lcdm\ halo
may continue growing hierarchically, specially in the case of more massive halos. 
Therefore, for a given present-day \ms, the earlier 
the galaxy is quenched (hence the redder it is), the lower tends to be its \ms-to-\mh\ ratio,
in spite that its SF efficiency could have been high when it was active. 
This means that the redder the galaxy, the lower is the \ms-to-\mh\ ratio, as we have found here.
Galaxy quenching is consistent with the observational fact that, on average, as more 
massive are the galaxies, the earlier they have formed and {\it ceased their SF} 
\citep[e.g.,][]{Thomas+2005,Bundy+2006,Bell+2007,Drory+2008,Pozzetti+2010}.
This phenomenon is known in the literature as ``archaeological" 
downsizing \citep[see][and more references therein]{Fontanot+2009,Conroy+2009,Firmani+2010a}.

\begin{figure}
\vspace*{-230pt}
\hspace*{10pt}
\includegraphics[height=6.5in,width=6.5in]{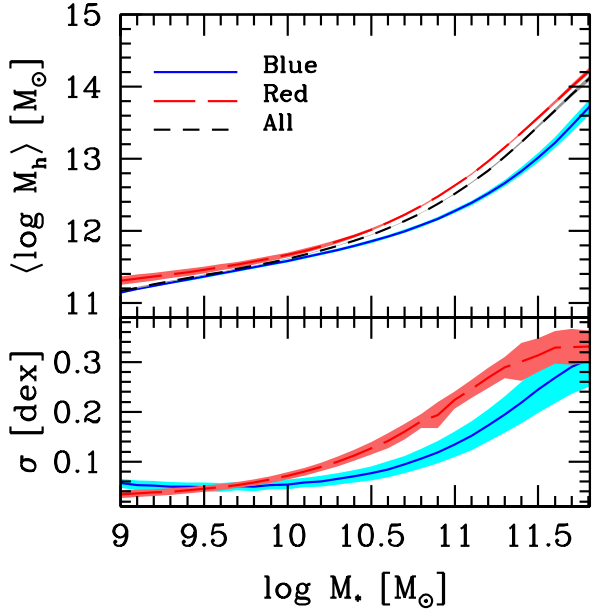}
\caption{ \textit{Upper panel:} 
Solid, long dashed and short dashed lines show the resulting inverse \shmr s from our best fitting model,
for blue, red and all central galaxies, respectively.
The shaded areas correspond to the confidence intervals from the MCMC trials. 
\textit{Lower panel:} Scatters around the inverse \shmr s for
blue and red central galaxies; the shaded areas correspond to the confidence intervals from the MCMC trials.
These scatters are upper limits to the intrinsic scatters.
}
\label{Mh-Ms}
\end{figure}

We calculate also the inverse \shmr s, that  is, dark matter halo mass as a function of        
stellar mass for blue, red and all central galaxies.  Inverting the \shmr\ is not just inverting the 
axises of this relation. Due to the scatter, there is a non-negligible change in the inverse 
slopes of the \shmr\ when passing to the halo-stellar mass relation, specially at high masses 
\citep[][]{Behroozi+2010}. We compute the mean \mh\ as a function of \ms\ for blue and red 
galaxies ($j=b$ and $r$, respectively) as;
\begin{equation}
	\langle\log(\mh)\rangle_j(\ms)=\frac{\int\Pcj\phi_{h,j}
	(\mh)\log\mh d\mh}{\int\Pcj\phi_{h,j}(\mh)d\mh},
\end{equation}
and their corresponding intrinsic scatter,
\begin{equation}
	\sigma_j(\ms)=\left(\frac{\int\Pcj\phi_{h,j}
	(\mh)\mu^2d\mh}{\int\Pcj\phi_{h,j}(\mh)d\mh}\right)^{1/2},
\end{equation}
where $\mu=\log\mh-\langle\log(\mh)\rangle_j(\ms)$.
In the upper panel of Fig. \ref{Mh-Ms}, we plot the \mh--\ms\ relation for the blue and 
red central galaxies. The shaded areas correspond to the 1$\sigma$ confidence levels from 
the MCMC trials. The obtained total scatters, $\sigma_b(\ms)$ and $\sigma_r(\ms)$, are 
shown in the lower panel; the shaded areas are the confidence levels. 
For blue galaxies, $\sigma_b(\ms)$ changes from $\sim0.06$ dex to $\sim0.31$ dex, 
while for red centrals $\sigma_r(\ms)$ changes from $\sim0.03$ dex to $\sim0.33$ dex. 
As in the direct \shmr s, these scatters are an upper limit to the intrinsic ones. 

As expected, the dark matter halo of red centrals is on average more massive than the 
one of blue centrals, specially for galaxies more massive than $\ms\sim 2\times 10^{10}$ \msun.
This is also consistent with the fact that red centrals are more clustered than
blue centrals of the same stellar mass (Fig. \ref{f2a}). Actually, it is rare to find a blue central
in group/cluster sized halos, but if this is the case, then its host halo is 
significantly (up to a factor of $3$) less massive than the one of a red central of the 
same stellar mass, and therefore, it is expected to be less clustered. 
On the side of low-mass galaxies, it is rare to find red galaxies, but if this is the case,
they also have (slightly) more massive halos than the blue ones, or less stellar masses
for a given \mh. This can be due to strong early SN-driven outflows leaving these rare galaxies
devoid of gas and in process of aging (reddening), while, actually most of low-mass
galaxies instead seem to have delayed their SF histories, not suffering then strong SF-driven outflows
and being blue today.

\subsection{Scatter and color segregation in the \shmr}
\label{scatter-discussion}

The scatter around the average (total) \shmr\ has been discussed previously in the literature 
\citep[see e.g.,][and more references therein]{Yang+2008,Cacciato+2009,More+2011, 
Li+2012,Leauthaud+2012, Reddick+2013,RAD13,Kravtsov+2014}. 
Typically, these previous works constrained the scatter around the total \shmr, 
rather than constraining separately $\sigma_b$ and $\sigma_r$ \citep[but see][]{More+2011}.
In other words, they {\it assumed} that the scatter around the \shmr\ is an unimodal 
(lognormal) and random distribution.
Moreover, in these works it is assumed that $\sigma_A$ is {\it constant}, with reported values of 
$0.15-0.20$ dex, which are close to our value of $\sigma_r$ at large masses. 
In fact, this is expected since the majority of halos more massive than 
$\mh\sim3\times 10^{12}$ \msun\ host 
red centrals, and these are namely the masses explored in most of the cited works.  
To illustrate this point we compare the scatter constrained by some of these works
in Fig. \ref{fig-scatter}.\footnote{Recall that our constrained scatters are    
actually upper limits to the intrinsic scatters since they are convolved with 
the measurement errors (see Section \ref{scatter}).}           
We have also estimated $\sigma_r$ and $\sigma_b$ from the             
\citetalias{Yang+2012} galaxy group catalog for seven different halo mass bins equally
spaced in a width of 0.3 dex and only for those above $\mh=5\times10^{12}\msun$. We
compute these scatters following \citet{Yang+2009b}, and find that  
$\sigma_r\sim0.14$ dex, in agreement with our result, while $\sigma_b\sim0.15$ dex, which is
slightly larger than our determination.

Our results show that the mean \shmr s  of blue and red central galaxies and    
the scatters around them are different. This implies that the distribution of all 
central galaxies is bimodal, as is shown in Fig. \ref{f11}, and that the color is 
one of the sources of the intrinsic scatter around the average \shmr. Note that if 
the \shmr s of blue and red centrals constrained with the observations were 
statistically similar, (i.e., the same probability distributions for blue and
red centrals), then this would mean that the \shmr\ of all centrals is not 
segregated by color (the scatter is not due to color). 
As mentioned above, for large masses, red centrals completely 
dominate in such a way that the average \shmr\ and its scatter are close to 
the mean \shmr\ and the scatter of red centrals, respectively (Fig. \ref{f3}).  
In this sense, the scatter distribution of the average \shmr\ is close to an 
unimodal distribution, the one of the red galaxies (see Fig. \ref{f11}).

At smaller masses, $\mh\lesssim 3\times 10^{12}$ \msun, the fractions of blue 
(red) centrals increases (decreases) with decreasing mass in such a way that 
the intrinsic scatter around the average \shmr\ of central galaxies is  
bimodal with a significant separation between the peaks (see Fig. \ref{f11}). 
According to Fig. \ref{fig-scatter}, its scatter increases with decreasing halo mass. 
This result implies that at masses where the fractions of blue and red central 
galaxies are not significantly different, {\it the use and interpretation of the average 
(total) \shmr\ should be taken with care}. 
If the study refers to the intermedium-low mass central galaxy population, 
the large intrinsic scatter around the \shmr\ (see Fig. \ref{fig-scatter}) should be taken into 
account for any inference. If the \shmr\ is used in studies where a distinction is 
made in between blue and red (late- and early-type) galaxies, then the \shmr\ 
separated into blue and red galaxies should be used, 
given the significant segregation by color that we have found in this
relation. For instance, this is the case of studies where the \shmr\ is connected with observable correlations
as the Tully-Fisher relation for late-type (blue) galaxies and the Faber-Jackson 
relation for early-type (red)
galaxies.

\subsection{Comparison with previous works}
\label{comparison}

\begin{figure}
\vspace*{-120pt}
\hspace*{10pt}
\includegraphics[height=5.5in,width=5.5in]{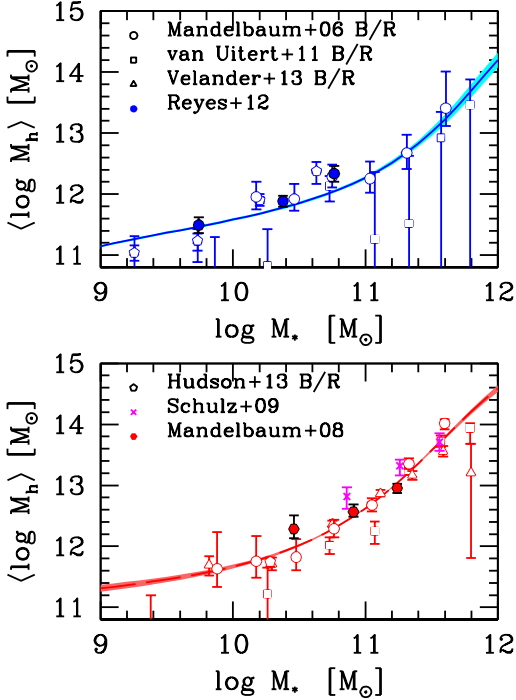}
\caption{ Halo mass as a function of stellar mass (inverse \shmr), 
similar to Fig. \ref{Mh-Ms}.  Model results are
compared with several galaxy-galaxy weak lensing studies
indicated inside the panels.
}
\label{weak-lensing}
\end{figure}

\begin{figure}
\vspace*{-120pt}
\hspace*{10pt}
\includegraphics[height=5.5in,width=5.5in]{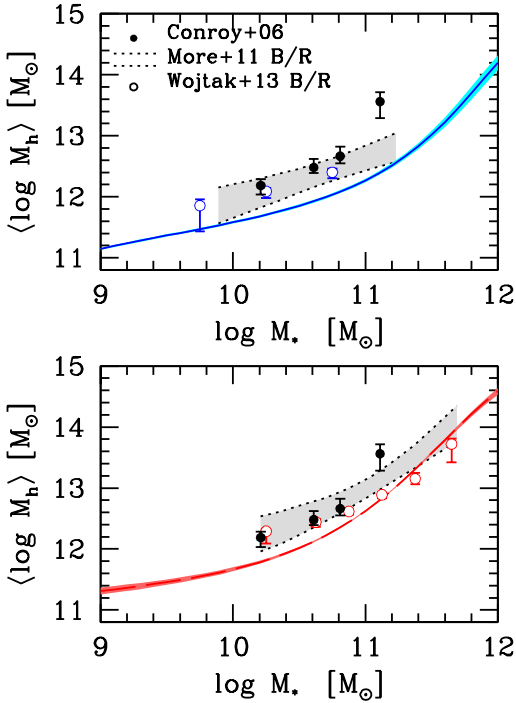}
\caption{Similar as Fig. \ref{weak-lensing} but this time comparing with several 
stacked satellite-kinematics studies indicated inside the panels.
}
\label{sat-kinematics}
\end{figure}

In this subsection, we compare our constrained \shmr s for local blue and 
red central galaxies with those previously obtained using direct methods,
namely galaxy-galaxy weak lensing and satellite kinematics. We also compare
our average \shmr\ with those of previous semi-empirical studies.
Where necessary, we apply corrections to the stellar 
mass reported by different authors to be consistent with 
the \citet{Chabrier2003} IMF adopted here
(see e.g., table 2 in \citealp{Bernardi+2010}). 
When necessary, we also correct the halo masses to match our definition
of virial mass. The corrections are done according to the relations
reported in Appendix \ref{halo-mass}.

\subsubsection{Comparisons with direct methods}  

First, we compare our results with those obtained from {\it galaxy-galaxy
weak lensing}. In these studies, in order to attain an acceptable signal-to-noise,
observations of individual galaxies are stacked in bins of \ms\ (or luminosity). 
Therefore, these measurements refer to halo mass as a function of \ms. In the upper 
and lower panels of Fig. \ref{weak-lensing}, we  
reproduce our resulting $\langle\log\mh\rangle(\ms)$ 
relations for blue and red galaxies, respectively, as plotted in Fig. \ref{Mh-Ms}, and we
compare them with several weak-lensing results. Note that in this case shaded areas represent
the uncertainties around the $\shmr$s.   
\citet[][empty blue circles, error bars are the 95\% confidence intervals]{Mandelbaum+2006} 
used SDSS DR4 galaxies separated into late- and early-type galaxies according to the 
bulge-to-total ratio as given by the parameter frac\_deV provided in the SDSS PHOTO 
pipeline (a de Vaucouleours/exponential decomposition was applied). 
\citet[empty blue squares][]{vanUitert+2011}
used the combined image data from the Red Sequence 
Cluster Survey (RCS2) and the SDSS DR7
to obtain the halo masses for late- and early-type galaxies as a function of \ms. 
(they also used the  frac\_deV parameter for defining the morphology). 
In the figure, we included measurements over only the redshift range $z=[0.08,0.41]$.
Both \citet[][empty blue triangles]{Velander+2014} and 
\citet[][empty blue pentagons]{Hudson+2014} used the Canada-France-Hawaii Telescope 
Lensing Survey to derive halo masses of blue and red galaxies; the latter division was done
based on the bimodality in the color--magnitude diagram. Halo mass measurements 
derived in \citet{Velander+2014} are on average at redshift $z\sim0.3$, while for \citet{Hudson+2014}
we have plotted only those measurements below $z=0.31$.
We also plot the results from \citet{Mandelbaum+2008} (red solid circles)
and \citet{Schulz+2010} (magenta crosses)  for massive central 
early-type galaxies based on the local SDSS DR7 sample
and a more sophisticated criteria for selecting early-type lens population. 
 
Our \shmr\ determinations are consistent, within the uncertainties, with the various 
weak-lensing studies, which cover each one different mass ranges and have large uncertainties.  
As mentioned in Section \ref{results}, a source of discrepancy between different authors is
due to the way blue and red (or late- and early-type) galaxies have been defined. 
In addition, results on $\langle\log\mh\rangle$ as a function of \ms\ 
are sensitive to the fact that different 
surveys may have different levels of measurement error in their stellar mass
estimates, particularly those based on photometric surveys, \citep[e.g.,][]{Leauthaud+2012}.              
In spite of all of these caveats, the overall agreement between these previous studies
with our results is encouraging. 
 
In Fig. \ref{sat-kinematics}, we compare our inverse $\shmr$s of blue and red centrals
(as in Fig. \ref{weak-lensing}) with stacked {\it satellite kinematics} studies. 
\citet[][shaded gray area]{More+2011} used the \citet{Yang+2007} group catalog and the
spectroscopic velocities of the SDSS survey.  \citet[][open circles]{Wojtak+2013} used
the SDSS DR7 spectroscopic catalog for blue and red central galaxies. We also plot
the results by \citet[][solid circles]{Conroy+2007}, though these authors did not present
their results for the \mh-\ms\ relation separated into blue and red galaxies 
(then, their data are repeated in the upper and lower panels); 
they combined data from the DEEP2 Galaxy Redshift Survey and the SDSS DR4 (their halo 
masses have been corrected by $\sim30\%$ due to incompleteness, see their Appendix A). 
As seen, the satellite kinematics method tends to give 
higher halo masses than our semi-empirical results and than weak-lensing studies.
The discrepancy between satellite kinematics and other methods has been noted previously, 
\citep[see e.g.,][]{More+2011,Skibba+2011,Rodriguez-Puebla+2011}. 
The differences can be partially explained by the relation between $\mh$ and the number of satellite
galaxies at a fixed $\ms$.                       
Since the technique is based on the
kinematics of satellites, these studies can be biased to higher halo masses due to the loss
of data in the case of those systems lacking satellites or with poor kinematical information,
namely those of smaller halo masses at a given stellar mass.

\begin{figure}
\vspace*{-260pt}
\hspace*{0pt}
\includegraphics[height=6.5in,width=5.8in]{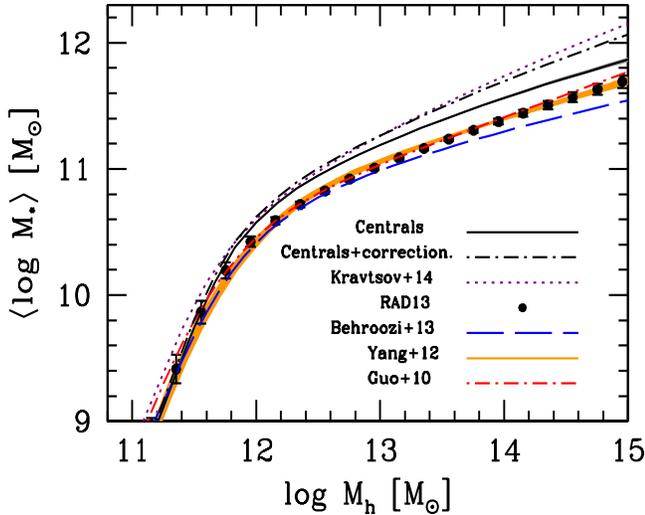}
\caption{ Average \shmr\ for all central
galaxies constrained by our method (see Eq.~\ref{msmh}). This is indicated with the
solid line surrounded by a gray shaded area which shows the model uncertainty around this relation. 
For comparison, the results obtained in \citet[][red long-dashed curve]{Guo+2010},
\citet[][blue dot-dashed curve]{Behroozi+2013}, 
\citetalias{Yang+2012}(orange shaded area indicate their 68\% of confidence), 
and \citetalias{RAD13} (dots with error bars) are reproduced.
In addition, we plot our resulting \shmr\ for all central galaxies that 
takes into account more adequate light profile fittings to galaxies (black long-dotted curve).
This is compared with the \shmr\ for all galaxies reported in
\citet[][violet dotted lines]{Kravtsov+2014}.
}
\label{f7}
\end{figure}

\begin{figure}
\vspace*{-245pt}
\hspace*{0pt}
\includegraphics[height=6.5in,width=6.5in]{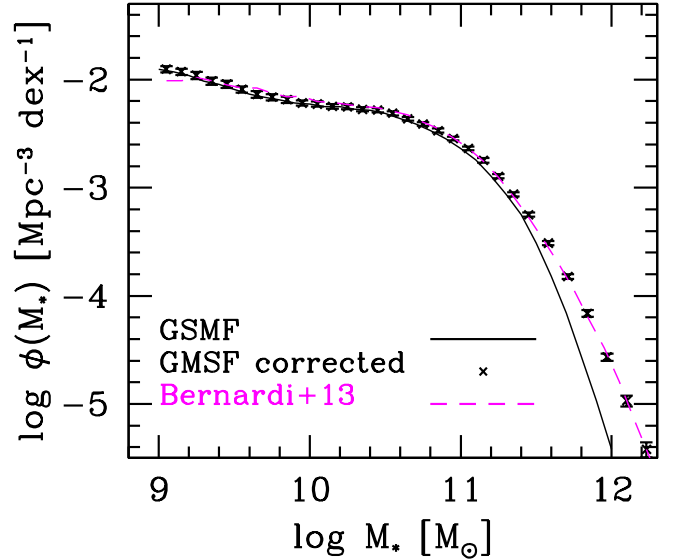}
\caption{\gsmf\ as inferred in Section \ref{data}  but introducing a correction on the stellar masses due to
new light profile fittings to galaxies, skeletal symbols with error bars. 
The magenta short dashed line line shows the \gsmf\ reported in 
\citet{Bernardi+2013}, who calculated stellar masses with the new light profile fittings (see text for details). 
For comparison, the solid line shows the \gsmf\ for all galaxies from the MPA-JHU \nyu/SDSS DR7 sample
obtained in Section \ref{data}.
}
\label{f7a}
\end{figure}

\subsubsection{Comparison with previous inferences of the average \shmr}
\label{mass-correction}

Figure~\ref{f7} compares the {\it average} \shmr\ of central galaxies (Eq.~\ref{msmh}) 
constrained by our method with those reported in \citet[][red long-dashed curve]{Guo+2010},
\citet[][blue dot-dashed curve]{Behroozi+2013}, \citetalias{Yang+2012} (orange shaded area 
indicate their 68\% of confidence), and \citetalias{RAD13} (dots with error bars). 
In the first two works,  
the \shmr\ was obtained by matching the abundances of all galaxies 
to the abundances of halos plus subhalos, therefore, it is rather the \shmr\ of all galaxies. 
Instead, in the case of the last works, 
their $\shmr$s are only for central galaxies (their set SMF2 and set C, respectively). 
The \shmr\ of central and all galaxies do not differ actually too much because the \shmr\ 
of satellite galaxies is close to the one of central galaxies (\citetalias{RAD13}).

At masses below $\mh\sim 10^{12}$ \msun, our average (blue + red galaxies) \shmr\ for centrals
is close to the $\shmr$s reported in the above cited studies. In contrast, at larger masses, 
our \shmr\ increases more rapidly than these studies.  
The main reason is that at large masses our calculation of the \gsmf\ falls slightly shallower 
than in previous works, see Fig.\ref{f1}.  However, it could be that 
the high-mass end of the \gsmf\ is even shallower than our determination!

Recently, several studies have pointed out to a systematical
underestimation of luminosity and stellar mass-to-light ratios 
of galaxies due to the commonly used aperture
limits in the SDSS \citep[see][and more references therein; 
see also Mendel et al. 2014]{Bernardi+2013}.
Surface brightness (mass) profiles of galaxies, in particular the central ones in clusters, 
extends much further away than the commonly used apertures 
\citep[][and more references therein]{Kravtsov+2014}.
In \citet{Bernardi+2013}, luminosity and stellar mass functions
were calculated based on different model fits for the surface brightness profiles in the SDSS 
galaxies. The authors showed that their results preferred
$\gsmf$s with the most luminous galaxies having larger masses 
for a given number density than most 
of the previous published \gsmf s.  Similar conclusion were obtained in \citet{He+2013} based
on a more sophisticated photometric data reduction from the SDSS DR7 and with morphological classifications 
from the Galaxy Zoo project \citep{Lintott+2011}. 
\citet{Kravtsov+2014} have confirmed this by using 
a compilation of well studied massive central cluster galaxies. 
\citet{Mendel+2014} have compared the MPA-JHU DR7
stellar masses with the ones obtained by computing accurate 
bulge+disk and S\'ersic profile photometric
decompositions in several bands. In the case of S\'ersic profile, they find masses larger by
$\approx 0.08$ dex at the smallest masses and by $\approx 0.23$ dex at the largest masses.
Following their results, we correct conservatively our masses 
by $\approx 0.05$ dex for masses up to 
log(\ms/\msun)$\sim 10.7$ and then increase smoothly the correction ending with 
0.23 dex at log(\ms/\msun)$\sim 12$. 

In Fig. \ref{f7a}, we reproduce the resulting total $\gsmf$ by correcting stellar masses 
as described above (skeletal symbols with error bars). 
For comparison, in the same figure we include 
the \gsmf\ for all galaxies from the MPA-JHU \nyu/SDSS DR7 sample
obtained in Section \ref{data}, solid line. The corrected 
$\gsmf$ is consistent with previous estimates, except at the
high-mass end, which has a significantly shallower fall than most of previous ones, 
but in good agreement with \citet[][their S\'ersic profile
case for the \ms\ estimate, magenta short dashed line]{Bernardi+2013}.
These authors extensively discuss about how
sensitive is the mass determination of the most luminous galaxies on the way the
light profile is fitted. The spirit of the correction introduced above to our stellar masses 
was namely to take into account more adequate light profile fittings to 
galaxies, specially the most massive ones, as was done in  \citet{Bernardi+2013}.

In Figure \ref{f7} we plot the resulting \shmr\ for all central galaxies 
when using the corrected stellar masses (black long-dotted curve).
For comparison we also reproduce the \shmr\ for all galaxies reported in
\citet[][violet dotted lines]{Kravtsov+2014}. Both results are similar and they show that
when taking into account more adequate light-profile fittings to galaxies, specially the most massive 
ones, the \shmr\ increases more rapidly than previous reports.
However, we highlight that all the results presented in previous sections we used a \gsmf\ 
estimated based on the SDSS standard light-profile fittings, see Section \ref{gsmfs-obs}.

\section{Summary and Conclusions}
\label{conclusions}

By means of a semi-empirical galaxy-halo connection model, we
have inferred the \shmr s of local {\it blue and red} central galaxies as well as
their intrinsic scatters. The \shmr\ of {\it all} central galaxies is the density 
average of these \shmr s. Our parametric model is a combination of 
the \amt, HOD model and CSMF formalism. The model allows us to separate 
the fraction of halos hosting blue/red central galaxies at each halo mass. 
The parameters of the model were constrained by using the
\gsmf s of blue and red galaxies inferred here from the SDSS DR7, 
divided into central and satellite components according to the 
\citetalias{Yang+2012} galaxy group catalog,
and the correlation functions of blue and red galaxies in different \ms\ bins \citep{Li+2006}. 
The criterion of the latter authors, based on the color--magnitude diagram, 
is used to separate the samples into blue and red galaxies. The main results 
obtained with our semi-empirical approach are as follows:

$\bullet$ The mean \shmr\ of blue and red central galaxies are different at a significant 
statistical level. At a given \mh, blue centrals have larger \ms\ than red centrals. 
At log(\mh/\msun)$\approx 11.7$, the difference attains its minimum, $0.16$ dex.
At larger masses, it increases up to $0.24$ dex for log(\mh/\msun)$\approx 12.7$, remaining 
then roughly constant. At smaller masses, the difference strongly increases.
These differences are larger than the $1\sigma$ confidence levels of each relation
and their corresponding constrained scatters. The \ms-to-\mh\ ratio of blue (red) centrals peaks at 
log(\mh/\msun)$\approx 12.17$ ($\approx 12$) with a mean value of 0.051  (0.031). 

$\bullet$  The density-averaged \shmr\ for all central galaxies lies in between the \shmr\ of blue 
and red galaxies, but closer to the former at log(\mh/\msun)$< 11.5$ and closer
to the latter at log(\mh/\msun)$>12.5$. This is because blue and red central galaxies
dominate below and above these masses, respectively. At the mass interval 
$11.5\lesssim$ log(\mh/\msun)$\lesssim 12.5$, the conditional stellar mass distribution (scatter)  
of central galaxies is strongly bimodal and color dependent. 

$\bullet$ The constrained scatters around the blue/red/average \shmr s are small.     
The width of the assumed lognormal function for the conditional stellar mass distribution (scatter) 
is slightly smaller for blue centrals than for red centrals: 
$\sigma_b\approx0.12$ dex and $\sigma_r\approx0.14$ dex, respectively. 
Note that the values constrained for these scatters are composed by an 
intrinsic component and by a measurement error component.

$\bullet$ The scatter of the average \shmr\ changes from $0.20$ dex 
to $0.14$ dex for $\log(\mh/\msun)\sim11.3$ to $\log(\mh/\msun)\sim15$, respectively
(the intrinsic scatter component is expected to be smaller). 
The increasing towards lower masses is due to the color bimodality in the conditional 
\ms\ distribution at these masses.
In previous studies, the scatter for all central galaxies has been assumed constant.

$\bullet$ The model predicts other distributions of the galaxy central and satellite populations, 
for both blue and red galaxies, which agree with independent observational determinations  
Among them, we remark: 
\begin{enumerate}
\item The dependence of the blue/red central galaxy fractions on \mh. We have assumed
a functionality for this dependence based on observational studies, and with our 
method the free-parameters of this functionality were constrained.  
At log(\mh/\msun)$=11$, 
around $\approx87\%$ of centrals are blue; this fraction decreases with \mh; at 
log(\mh/\msun)$=11.83$ half of centrals are blue and half are red; 
for group masses, log(\mh/\msun)$>13$, the centrals are red in more 
than 90\% of the cases. 
These results agree with weak lensing determinations. 

\item The satellite population is dominated by blue galaxies in low mass halos, where
blue galaxies also dominate among the central population. In contrast, red satellites
dominate in massive halos, where red galaxies also dominate among centrals. The 
predicted satellite \cmf s for $\mh\grtsim 10^{13}$ \msun\ agree well with those from 
the observational galaxy group catalog of \citetalias{Yang+2012}  \citep[see also][]{Yang+2009b}. 

\end{enumerate}

Our findings show that blue central galaxies have higher \ms-to-\mh\ ratios than red centrals. 
However, this does not mean that the former have more efficient SF rate histories than the latter.  
Instead, this can be interpreted as that red centrals are such because they halted 
their \ms\ growth in the past, likely by SF quenching processes. Such an interpretation is better
seen in the inverse \shmr, which shows that for a given \ms, red centrals reside in more 
massive halos than blue centrals.  This is likely because the stellar mass growth of red 
centrals is halted (mainly due to quenching) while 
their halos continue growing hierarchically. Since the difference in \mh\ between red and
blue galaxies increases with \ms, the quenching redshift is expected to happen earlier 
as the galaxy is more massive (downsizing). 
That red galaxies have on average more massive halos than blue one at a given \ms,
implies also that the former are more clustered than the latter, as observations show.

\acknowledgments 
A. R. acknowledges Shanghai Jiao Tong University (SJTU) postdoctoral fellowship.  
V.~A.\ acknowledges CONACyT (ciencia b\'asica) grant 167332.
V.~A. thanks the hospitality of the Center for Astronomy and Astrophysics
of SJTU,  where this paper has been finished.
This work is supported by the grants from NSFC (Nos.  11121062,
11233005) and by the Strategic Priority Research Program ``The
Emergence of Cosmological Structures" of the Chinese Academy of
Sciences, Grant No. XDB09000000. 
Part of the material of this paper was presented in the Ph.D. Thesis (UNAM) by A. R.  

We thank to Ramin Skibba and Emmanouil Papastergis 
for their comments on an earlier draft as well as for detecting typographical errors. 
We also thank to Ivan Lacerna for useful discussion about this draft. 
We thank Mariangela Bernardi for providing us in 
electronic form her data for the \gsmf\ as well as 
Surhud More and Ramin Skibba  for their fraction of blue central galaxies as a function of halo mass. 

\bibliographystyle{mn2efix.bst}
\bibliography{Bibliography}

 \appendix
 \section{A. K-correction}
\begin{figure*}
\includegraphics[height=2.3in,width=3.6in]{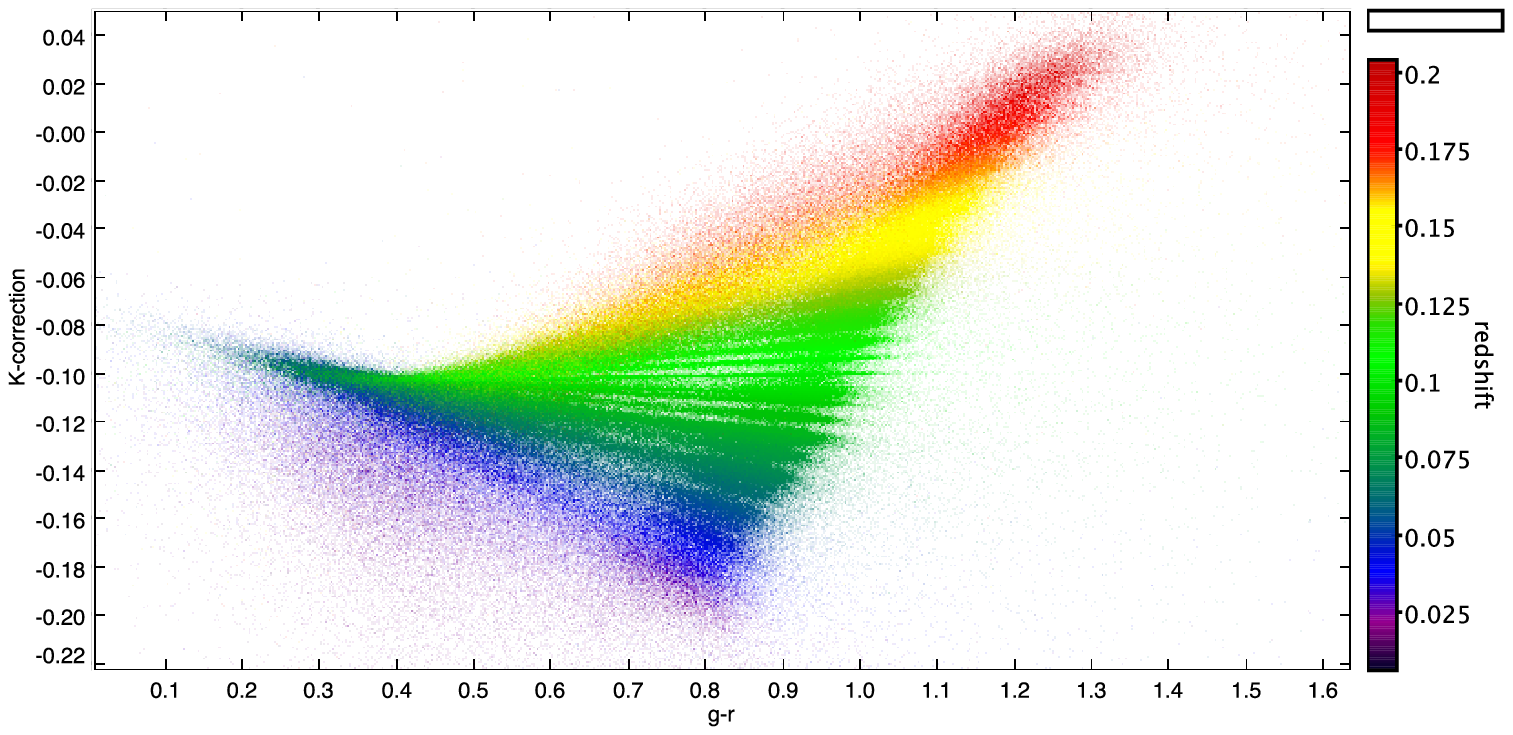}
\includegraphics[height=2.3in,width=3.6in]{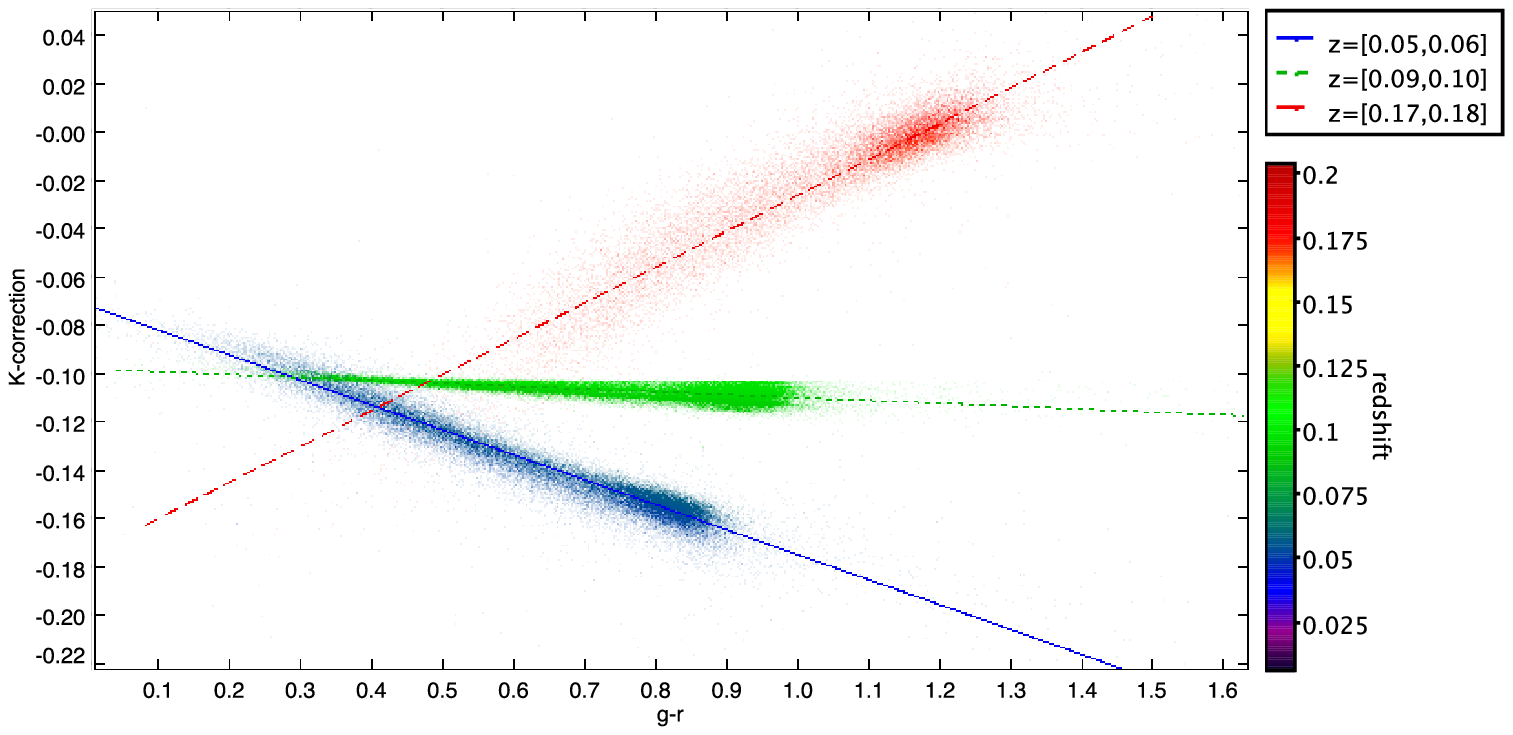}
\caption{\textit{Left panel:} K-correction from the NYU-VAGC \citep{Blanton+2005} 
as a function of $(g-r)$ color and redshift. \textit{Right panel:} The same as left panel but
for three different redshifts bins. The different lines show the best fits for a relation of the form
$K(g-r)=m\times(g-r)-b$ for these redshifts bins. 
}
\label{Kcorr}
\end{figure*}

In this Appendix, we describe our analytical model for the K-correction. 
Figure \ref{Kcorr} shows the K-correction (to $z=0.1$) as a function of $(g-r)$ colors (to
$z=0.0$) and redshift. Recall that our redshift range is $0.01\leq z\leq0.20$. 
For this plot, we use the values of the K-corrections reported in the  
NYU-VAGC \citep{Blanton+2005}\footnote{Available at http://sdss.physics.nyu.edu/vagc/.} 
based on the SDSS DR7 and calculated for each galaxy from the 
{\it k-correction} program v$4\_1\_4$ \citep{Blanton+2007b}. As can be seen in right hand
panel, for a given interval of redshift the relation between the K-correction and color is very well 
approximated to a linear relation, i.e.,
$K(g-r)=a\times(g-r)-b$. 
By dividing our redshift range into 20 redshift bins of width 0.01, we have found
that the color and redshift dependence of the K-correction to $z=0.1$ is
\begin{equation}
		K(z,g-r)=a(z)\times(g-r)-b(z);
	\end{equation}
where
	\begin{equation}
		a(z)= \left\{ \begin{array}{rcl}
		 -1.649\log(1+z)-0.093 & {\rm for}   & z\leq0.045  \\
		 5.590\log(1+z)-0.231 & {\rm for}   & z>0.045,  \\
	\end{array}
	\right.
\end{equation}
and
\begin{equation}
	b(z)= \left\{ \begin{array}{rcl}
	 6.276\log(1+z)-0.181 & {\rm for}   & z\leq0.045  \\
	 -2.303\log(1+z)-0.017 & {\rm for}   & z>0.045.  \\
	\end{array}
	\right.
\end{equation}

The above approximation recovers, on average, to $\sim10\%$, and $\sim1\%$ the values of the
K-correction and absolute magnitudes, respectively.

\section{B. Halo mass transformations}
\label{halo-mass}

In order to find the differences in halo mass depending on the different definitions of it,
we employed halo catalogs from the Bolshoi simulation
\citep{Klypin+2011}, where halos are identified by means of the Bound-Density Maxima halo
finder algorithm \citep{Klypin+1997}. The advantage of using these catalogs is that they report
masses based on different definitions of halo mass, so we can use them in order to obtain
average correlations between these mass definitions.  
To do so, we use a sample of $\sim94,000$ halos. We find the following relations
for the virial halo mass used here ($\mh\equiv M_{\Delta_{\rm vir}}$):
\begin{equation}
	\log (M_{\Delta_{\rm vir}}/\msun)=1.014\log(M_{\Delta_m}/\msun)-0.07,
		\label{m200}
\end{equation}
and 
\begin{equation}
	\log (M_{\Delta_{\rm vir}}/\msun)=0.98\log(M_{\Delta_{\rm crit}}/\msun)+0.21,
		\label{crit200}
\end{equation}
where $\Delta_m=200$ and $\Delta_{\rm crit}=200$
are 200 times the mean matter and critical densities, respectively.

\section{C. Probability distributions}
\label{PD}
 
\begin{figure*}
\includegraphics[height=7in,width=7in]{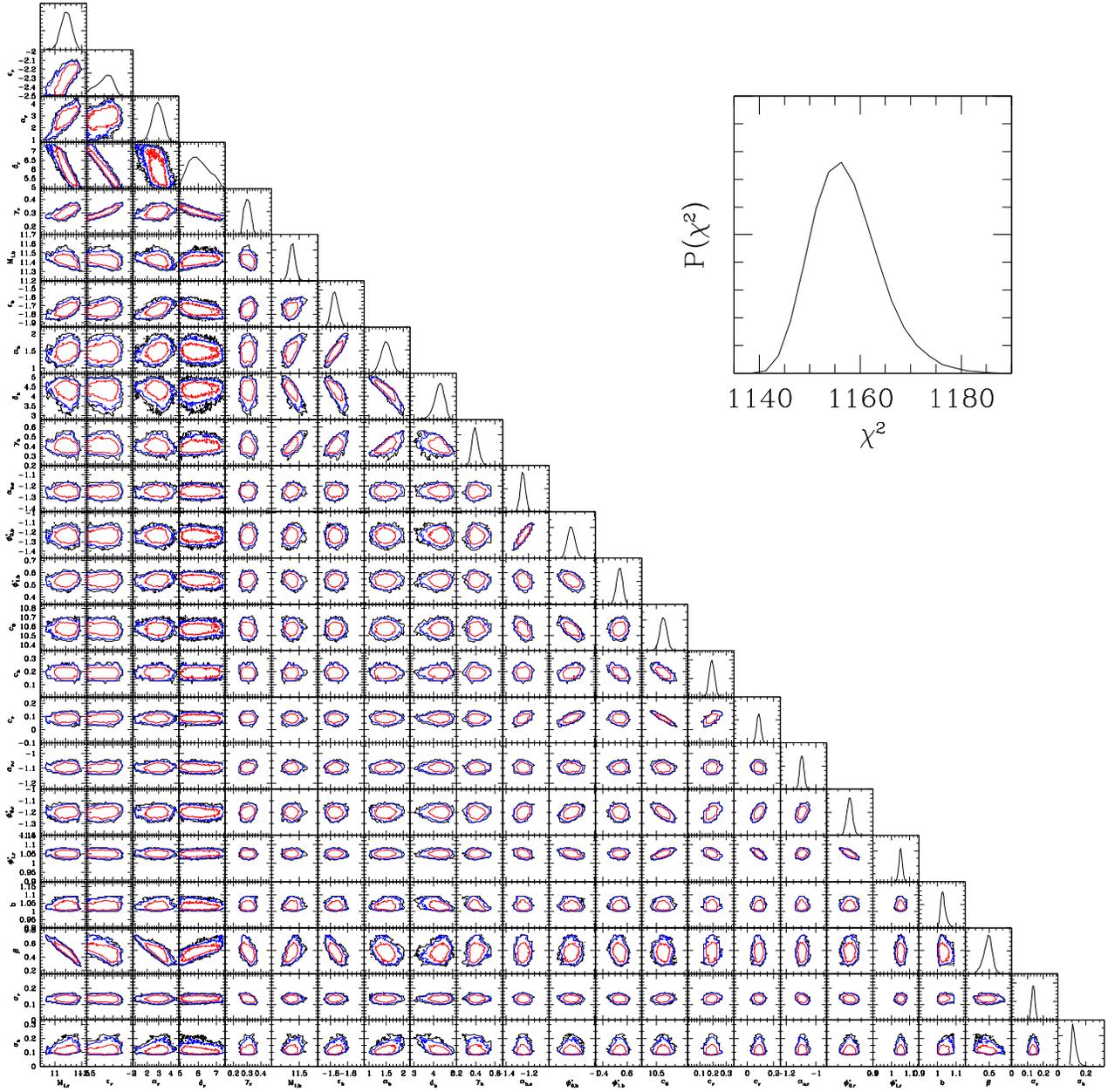}
\caption{Posterior probability distributions of our model parameters. Black contours
represent the $90\%$ of the models with the lowest $\chi^2$, while blue and red contours are
the same but for $68\%$ and $10\%$ of the models.
}
\label{PD1}
\end{figure*}

In Fig. \ref{PD1} we show the 1D (diagonal) and 2D posterior distributions of the model parameters
as constrained in Section \ref{results}.   Note that in most of cases the constrained parameters
do not correlate significantly among them.

\end{document}